\documentclass[preprint]{aastex}

\shorttitle{How Well Do We Know the Orbits of the Outer Planets?}
\shortauthors{Page, Wallin, \& Dixon}
\slugcomment{Submitted to The Astrophysical Journal, November 2008}

\usepackage{natbib}               % added by GLP 10-23-08
\begin{document}

\title{How Well Do We Know The Orbits Of The Outer Planets?}

\author{Gary L. Page\altaffilmark{1}, John F. Wallin\altaffilmark{2}, \and% 
        David S. Dixon\altaffilmark{3}}

%% \affil{George Mason University, 4400 University Drive, MS 5C3, Fairfax, VA 22030}
%% \email{gpage@gmu.edu}
%\author{David S. Dixon\altaffilmark{2}}
%% \affil{Jornada Observatory, Las Cruces, NM}
%\and
%\author{John F. Wallin\altaffilmark{3}}
%% \affil{George Mason University, 4400 University Drive, MS 5C3, Fairfax, VA 22030}

\altaffiltext{1}{George Mason University, Department of Computational and Data Sciences,% 
                 4400 University Drive, MS 6A2, Fairfax, VA 22030; gpage@gmu.edu.}
\altaffiltext{2}{George Mason University, Department of Computational and Data Sciences,% 
                 Department of Physics and Astronomy, 4400 University Drive, MS 6A2,% 
                 Fairfax, VA 22030; jwallin@gmu.edu.}
\altaffiltext{3}{Jornada Observatory, Las Cruces, NM; ddixon@cybermesa.com.}

\begin{abstract}
This paper deals with the problem of astrometric determination of the orbital elements of the 
outer planets, in particular by assessing the ability of astrometric observations to detect 
perturbations of the sort expected from the Pioneer effect or other small perturbations to 
gravity. We also show that while using simplified models of the dynamics can lead to some 
insights, one must be careful to not over-simplify the issues involved lest one be misled by 
the analysis onto false paths. Specifically, we show that the current ephemeris of Pluto does 
not preclude the existence of the Pioneer effect. We show that the orbit of Pluto is simply 
not well enough characterized at present to make such an assertion. A number of misunderstandings 
related to these topics have now propagated through the literature and have been used as a basis 
for drawing conclusions about the dynamics of the solar system. Thus, the objective of this 
paper is to address these issues. Finally, we offer some comments dealing with the complex 
topic of model selection and comparison.
\end{abstract}


\section{\label{intro}INTRODUCTION}

Recently, \citet{2006ApJ...642..606P} reported on the use of minor planets to assess 
gravity in the outer solar system. That paper was mostly devoted to the use of astrometry 
of asteroids to investigate gravity in the that region and the potential use of such 
observations to ascertain the reality of the Pioneer effect, an unexplained acceleration 
towards the Sun that perturbs the motion of Pioneer 10 and 11 beyond a distance of 20 AU 
\citep{2002PhRvD..65h2004A}. However, some statements were made that applied to comets 
and the outer planets. In particular, with respect to the outer planets, 
\citet{2006ApJ...642..606P} commented that the ephemerides of the outer planets are 
almost entirely based on optical observations and are much less accurate than those for 
the inner planets \citep{2004A&A...417.1165S}. This, coupled with the Pioneer effect not 
being observed in the inner solar system argues that the major planets are not good 
candidates for investigating gravity in the outer solar system.

\citet{2006NewA...11..600I}, carried out detailed calculations on the impact that the 
Pioneer effect would have on orbital motions of the outer planets. They found a Pioneer-like 
acceleration would produce significant secular and periodic effects and conclude that the 
absence of such evidence indicates that the Pioneer effect is ruled out as a phenomenon 
effecting the outer planets. The approach of these authors involves analytically and 
numerically investigating the Gaussian rate equations for rates of change in orbital 
elements as a function of time and a perturbing force. This approach will be discussed 
more fully later in Section \ref{results}. 

Additionally, \citet{2007PhRvD..76d2005T} observes that \citet{2006ApJ...642..606P} 
offered no calculations to substantiate its conclusions about the outer planets and 
indicates that other authors reach different conclusions about the use of the outer planets 
in this role. \citet{2007PhRvD..76d2005T} also sought to illuminate the issues involved by 
using a simplified four parameter model of orbital motion. Referencing several papers 
\citep{2006ApJ...642..606P,2006NewA...11..600I, 2007PhRvD..76d2005T}, \citet{2006MNRAS.370.1519S} 
calls the dispute about the observability of Pioneer effect-like perturbations in the motion of 
the outer planets indicative of a lack of consensus in discussions of the planet's motion.

Both \citet{2006NewA...11..600I} and \citet{2007PhRvD..76d2005T} refer to \citet{2005astro.ph..4634I} 
to buttress their contention that a Pioneer-like acceleration should be observable in the motion 
of the outer planets. In discussing options for a non-dedicated spacecraft investigation of the 
Pioneer effect, \citet{2005astro.ph..4634I} briefly considers whether the effect would be 
detectable in the motion of the outer planets and conclude that the Pioneer effect would have 
an impact on outer planet motion that would be at variance to that observed. This conclusion 
is largely based on the change in the $GM_{\odot}$ product that would be required to result in 
the Pioneer effect at the distance of Neptune and the comparison of that value with the otherwise 
known uncertainty of the $GM_{\odot}$ value. The basis of their conclusion seems to be similar to 
that expressed in \citet{1988PhRvL..61.1159T}, where modifications to Kepler's Law are used to 
assess the possibility of several modifications to gravity. \citet{1988PhRvL..61.1159T} restricts 
its attention to the case of the inner planets, however, where ranging data is available from both 
earth-based radar and spacecraft. However, \citet{2005astro.ph..4634I}'s conclusion that the 
Pioneer effect would have been observed in the motion of the outer planets is too all-encompassing. 
For example, some alternative theories of gravity provide for a variation in the force that a 
spacecraft would experience, as would distributed mass densities concentrated in the outer solar 
system. The issue here is whether such accelerations are really observationally detectable 
independent of their source.

This paper's purpose is twofold. First, we wish to deal with the problem of astrometric 
determination of the orbital elements of Pluto, in particular by assessing the ability of 
astrometric observations to detect perturbations of the sort expected from the Pioneer effect. 
Secondly, we wish to show that while using simplified models of the dynamics can lead to some 
insights, one must be careful to not over-simplify the issues involved lest one be misled by 
the analysis onto false paths.

Specifically, this paper shows that, contrary to recent assertions in the literature, the 
current ephemeris for Pluto does not preclude the existence of the Pioneer effect. We show 
that the orbit of Pluto is currently not well enough characterized to make such an assertion. 
Thus we address the views alluded to above that have now propagated through a number of papers 
(\citet{2006CQGra..23.7561J, 2006NewA...12..142M, 2006MNRAS.371..626S,% 
2007astro.ph..1848B,2007IJMPD..16.1475K, 2007EL.....7719001M, 2007arXiv0709.3690Z}). We believe 
that the conclusions presented in those papers are not established unambiguously and that caution 
needs to be used in drawing further inferences about the dynamics of the outer solar system.

Notwithstanding our comments about the dangers of using simplified methods, we note that our 
approach itself is a simplification. The outer planets must be dynamically assessed as a system. 
In order to rigorously determine whether a small perturbation like the Pioneer effect is detectable 
by astrometry, we should simultaneously include the changes it produces in the orbits of Uranus, 
Neptune, and Pluto. Only in this way can all the second order perturbations to the system be 
taken into account. However, in order to illustrate the ideas concerned and the weaknesses of 
the approaches outlined above, we restrict our attention here to manipulating the orbit of 
Pluto. This approach is further discussed in Section \ref{disc}.

The remainder of this paper is divided into four further sections. Section \ref{M&M} describes 
several ``back of the envelope'' approaches to understanding the nature of the orbital changes 
that a Pioneer effect would cause, and a detailed discussion of the analysis methodology we have 
chosen to make our case. Section \ref{results} addresses our results in terms of the relationships 
between orbital parameters and observation arc length on the observability of small perturbations. 
Finally, Section \ref{disc} provides a discussion of our results and Section \ref{concl} presents 
conclusions.

\section{\label{M&M}METHODOLOGY AND MODELS}

\subsection{Characterizing the Pioneer effect}
To begin, we must describe the working definition of the Pioneer effect used in this paper. 
Following \citet{1998PhRvL..81.2858A}, we take the Pioneer effect to be manifested by a radial 
acceleration, directed sunward, of magnitude $8.74\times10^{-10}$ m s$^{-2}$. Since the primary 
purpose of this paper is to investigate the dynamical consequences of the Pioneer effect, which 
apparently begin at about 20 AU from the Sun, and because there are no data indicating a more 
gradual onset of the Pioneer effect, we will assume that the anomalous acceleration begins 
abruptly at a heliocentric distance of 20 AU. 

We recognize that this is a simplistic model of the Pioneer effect. Alternative mechanisms exist 
that cause the acceleration to vary with object mass, orbital eccentricity, radial distance, and 
other parameters of the motion. As further observations are made of the Pioneer effect, they can 
be used to investigate different force models to explore various alternatives until the Pioneer 
effect is either ruled out or its origin is found. However, the current status of knowledge of 
the Pioneer effect argues that this simple model should be investigated first. Additionally, a 
perturbation beginning somewhat closer to the Sun at a more gradual pace would be more easily 
seen in the motion of the planets. As shown recently by \citet{2008AIPC..977..254S}, if the 
Pioneer effect occurred at shorter distances from the Sun, its effect on planetary ephemerides 
would have already been detected. Thus, our assumptions about the Pioneer effect represent the 
minimum plausible perturbation given the available data. Notwithstanding our concentration on 
the Pioneer effect, the analysis presented here should be valid for any small constant 
perturbation to gravity in the outer solar system. 

\subsection{Estimating Pioneer effect manifestations}
There are several ways to approach assessing whether small perturbations like the Pioneer effect 
can be observed in the motions of the outer planets. For clarity, however, let us make clear that 
we are talking of perturbations to the motion of the outer planets that are spherically symmetric 
and directed towards the Sun. Perturbations due to localized mass concentrations (i.e., Planet X) 
are specifically excluded. However, the latter can be instructive. Some years ago, there were 
allegations that the motion of the outer planets contained anomalies that indicated the presence 
of a large mass concentration in the outer solar system. \citet{1993AJ....105.2000S} showed how 
these presumed anomalies in Uranus' motion vanish when the orbital elements are adjusted while 
using correct values of Neptune's mass as determined by spacecraft. The important point here 
is that it is not enough to merely compare projected positions; rather, one must adjust the orbital 
elements, and even other parameters defining the problem, to best fit observational data. There 
are a number of approaches to addressing this problem.

One approach was taken by \citet{1988PhRvL..61.1159T} in which a variation in the value of the 
astronomical unit is investigated for the inner planets. Their analysis assumes a small 
eccentricity and makes use of ranging data available from radar and spacecraft observations. 
This shows that quite stringent constraints are placed on the nature of gravity at scales 
approximating that of the inner solar system. In particular, any Pioneer-like acceleration 
at those scales would long since have been detected.

Another approach is to consider the relative magnitudes of the acceleration due to the Sun 
and that due to the Pioneer effect. This is the approach taken in \citet{2006ApJ...642..606P}, 
although it was not documented in that paper. Certainly the Pioneer effect should not be 
expected to have a dominating impact on the motion of the outer planets. The ratio of the 
Pioneer acceleration to that produced by the Sun at a distance equal to the semimajor axis 
of the outer planets is 0.005, 0.013, and 0.023 percent for Uranus, Neptune, and Pluto, 
respectively. If we integrate an orbit without any perturbations to Newtonian gravity other 
than the Pioneer effect, and compare with the Keplerian case, we find the orbital periods of 
these objects are systematically shorter. Uranus' period shortens by 5.8 days and Neptune's 
by 24.1 days, while Pluto's period drops by 79.7 days. These intervals correspond to 0.02, 
0.04, and 0.09 percent of the periods of Uranus, Neptune, and Pluto, respectively.

Differentiating Kepler's Third Law implicitly, we obtain a relationship between a small 
change in an orbiting object's period and a corresponding change in the orbit's semimajor 
axis. In particular, $da/a = (2/3) dT/T$ where $a$ is the semimajor axis (in AU) and $T$ is 
the orbital period (in years). For Uranus, the fractional period shortening due to the Pioneer 
effect is equivalent to a fractional change in semimajor axis of approximately 0.013 percent. 
Similarly, the orbital period shortening is equivalent to a reduction in semimajor axis of 
0.027 and 0.059 percent for Neptune and Pluto, respectively.

These simple calculations imply an equivalent change in aphelion distance of $3.8\times10^{5}$, 
$1.2\times10^{6}$, and $4.3\times10^{6}$ km for Uranus, Neptune, and Pluto. In the cases of 
Uranus and Neptune, this is less than the approximate maximum errors in range of $2\times10^{6}$ 
km \citep{1992exaa.book.....S}). In the case of Pluto, the change in semimajor axis that would 
correspond to the shortening of its orbital period with the Pioneer effect is about twice 
the radial distance uncertainty. However, Pluto has completed less than one-third of an 
orbit since its discovery and its orbital elements are even less well-determined than the 
other outer planets. Thus, small changes in other orbital elements could easily obscure any 
orbital changes due to the Pioneer effect. 

\subsection{Celestial mechanics}
The values of orbital elements are not directly observable and to reduce observational data 
to orbital elements it is necessary to proceed by using numerical approximations\footnote{In 
this discussion we are dealing with ``classical'' observations consisting of two angles describing 
the position of the object in the sky at a moment in time. Orbital determination when one has range 
information, for example with radar observations, is a completely different mathematical problem. 
This latter area is sometimes called ``astrodynamics.'' Radar observations for objects subject to 
the Pioneer effect are technologically out of the question for the foreseeable future.}. The normal 
manner in which one proceeds is to first determine a ``preliminary orbit'' from a small number 
of observations and then to refine it by successive approximations into a ``definitive orbit'' 
as many more observations are added. 

The method used to improve the orbit and obtain the definitive orbit through additional 
observations is called ``differential correction'' and is well described in standard celestial 
mechanics texts (e.g., \citet{1914QB351.M92......}, \citet{1961mcm..book.....B}, or 
\citet{1988fcm..book.....D}). Differential correction uses a least squares approach to 
iteratively refine the estimates of the elements as more observations become available. 
Additionally, statistical information on the errors of the elements naturally results from 
the differential correction process. The process results in a set of orbital elements, along 
with error estimates for the elements and covariance parameters showing the degree to which 
the elements are correlated. It is worth remembering that the equations governing celestial 
mechanics are nonlinear and normally a linearized version of the problem is used to determine 
the covariances and elements. However, the nonlinearity manifests itself in occasional 
difficulties in determining elements and some of these difficulties will be discussed 
later in Section \ref{sect3-B}.

The whole chain of analysis outlined above for orbital element determination is associated 
with a number of errors that must be understood and characterized. Some are associated with 
the description of the solar system dynamics and calculation of the orbits, while some 
originate in the observations themselves. 

However, in determining elements from observations, there is no prior warning about the 
pathological situations that can arise. For example, if the inclination is small, the error 
in the longitude of the ascending node will be large. Similarly, if the eccentricity is small, 
the error in the argument of perihelion will be large and the time of perihelion passage 
or mean anomaly will be poorly defined. Such problems can be avoided through the use of 
alternative orbital elements (e.g., equinoctial elements).

Thus, the orbital characterization process outlined above must be carried out to enable 
predictions of the future positions of objects on the sky. The errors associated with the 
predictions must be compared statistically to determine whether a truly observable difference 
can be asserted between the alternative models of gravity.

Importantly, since orbital elements are not directly observable, it is not enough to simply 
integrate an object's equations of motion forward in time from a set of initial conditions, 
with and without the perturbation (as was done in \citet{2006NewA...11..600I}), and then 
compare the differences. One must first determine element values by adjusting them to a 
set of observations and a gravity model to obtain a new set of elements with associated 
error metrics. This can be understood physically by noting that the orbital elements 
describe the conserved mechanical energy of the body and if the potential field changes 
due to the existence or absence of a perturbation the elements need to be redetermined. 

In order to address these issues, we use the OrbFit\footnote{The OrbFit software and 
documentation are available freely for downloading from http://newton.dm.unipi.it/orbfit.} 
software package \citep{1999Icar..137..269M}. This program uses observational data and 
data on the dynamics of the solar system to determine orbital elements and predict 
ephemerides for minor bodies. Since OrbFit is available in source code, we added a simple 
option to include forces arising from the Pioneer effect. By comparing orbits resulting 
from synthetic observations with and without the Pioneer effect, we explore the expected 
effects of such a perturbation on asteroid orbits and examine when this effect can be 
detected astrometrically.

OrbFit is a complex piece of software that performs a number of tasks that relate to this 
analysis. The first task of interest here is that of orbit determination. This process 
begins with astrometric observations of the body of interest. An initial orbit is determined 
from a few of the available observations and a process of differential correction is applied 
to correct the initial elements and take into account any additional observations. 

The second task of interest performed by OrbFit is orbital position prediction. To predict the 
position of bodies of interest, OrbFit takes the position and velocity of the body of interest, 
and integrates the equations of motion of that body with many additional perturbations due to the 
other bodies in the solar system, solar oblateness, etc. 

Additionally, the process used by OrbFit naturally provides measures of the errors in elements that 
result from the unavoidable errors in position measurement and the associated element covariances. 
These data can be mapped on to positions in the sky and the error associated with the prediction, 
which permit our analysis to take place.

By using OrbFit for our calculations, we are able to include planetary and other perturbations and 
factors that affect the determination and prediction of orbits. These factors are dealt with in 
OrbFit to create orbital predictions and also conduct orbital analyses in a manner that very closely 
mimics the process by which data from observations are reduced to orbital predictions and elements 
in the real world. In particular, OrbFit is able to deal with various errors and noise that are both 
unavoidable and necessary for calculations at a level of accuracy sufficient for distinguishing 
between orbits affected by the Pioneer effect and those subject to gravity without the additional 
perturbation. 

The calculations outlined above represent standard techniques of celestial mechanics and can be 
performed by any number of systems. OrbFit was chosen because of its availability in source code. 
This permitted addition of a simple option to include the force arising from the Pioneer effect, 
which were simply added to the forces exerted on the bodies of interest by the major planets.

Rather than integrating the orbits of the main bodies in the solar system, their dynamics are 
introduced via the JPL DE405\footnote{The ephemeris files are available from 
http://ssd.jpl.nasa.gov/?ephemerides\#planets.} ephemeris. In this context the planetary ephemeris 
is a sort of lookup table containing positions of the planets so that their effect on the motion of 
a body of interest might be determined.

\subsection{Simulation of observations}
In this paper, we will parametrically vary the eccentricity of hypothetical test bodies, generate 
synthetic observations for them with-- and without a Pioneer-like acceleration, and determine the 
conditions under which the Pioneer effect can be observed. With the exception of eccentricity, the 
orbital elements of our test bodies are those of Pluto \citep{1992exaa.book.....S}. The eccentricity 
of the bodies are 0.001, 0.005, 0.01, 0.05, 0.1, 0.2, and 0.3. These values are chosen to bracket 
the actual eccentricity of Pluto while extending to nearly circular orbits similar to those found 
in the other outer planets. The nominal elements at the epoch J2000 (JD 2,451,545.0), are referred 
to the mean ecliptic and equinox of JD 2000.0 and are given in Table \ref{elements}.

\placetable{elements}
%\clearpage
\begin{deluxetable}{ll}
\tabletypesize{\scriptsize}
\tablecaption{\label{elements}Elements for the hypothetical bodies used in the analysis. The elements 
are at the epoch J2000 (JD 2,451,545.0) and are given with respect to the mean ecliptic and equinox 
of J2000. With the exception of the eccentricity, all elements have the values appropriate for Pluto.}
\tablewidth{0pt}
\tablehead{ \colhead{Element} & \colhead{Value\tablenotemark{a}} }
\startdata
Semimajor axis & 39.48168677 AU \\
Eccentricity & See text \\
Inclination & 17.14175 degrees \\
Longitude of Ascending Node & 110.30347 degrees \\
Argument of Pericenter\tablenotemark{b} & 113.76329 degrees \\
Mean anomaly\tablenotemark{c} & 14.86205 degrees \\
\enddata
\tablenotetext{a}{Source for element values is \cite{1992exaa.book.....S}}
\tablenotetext{b}{The argument of the pericenter is equal to the difference between 
the longitude of pericenter and the longitude of the ascending node.}
\tablenotetext{c}{The mean anomaly is the mean longitude minus the longitude of pericenter.} 
\end{deluxetable}

The observation cadence we have chosen to use roughly reflects that which exists for Pluto. If one 
reviews the observations for Pluto that were used to construct the DE405 ephemeris, we find 
approximate observation frequencies as indicated in Table \ref{obsfreqs}. With the exception of 
the 1914-1929 period, we have adopted these cadences and generate synthetic observations accordingly. 
The pre-discovery observations occuring from 1914-1929 are aggregated and an overall rate of one 
observation per year is used for this time period. 

\placetable{obsfreqs}
%\clearpage
\begin{deluxetable}{lrrl}
\tabletypesize{\scriptsize}
\tablecaption{\label{obsfreqs}Frequency of archive observations of Pluto used to develop recent JPL 
ephemerides. The earliest pre-discovery image of Pluto occurred in 1914. Observations after 2006 
are estimated (see text).}
\tablewidth{0pt}
\tablehead{ \colhead{Dates} & \colhead{No. Observations} & \colhead{No. Years} & \colhead{Approx. Cadence} }
\startdata
1914-1919 &   11 &   6 & 2 per year \\
1920-1929 &    6 &  10 & 1 per 2 years \\
1930-1939 &  431 &  10 & 1 per week \\
1940-1949 &  233 &  10 & 2 per month \\
1950-1959 &  113 &  10 & 1 per month \\
1960-1969 &  113 &  10 & 1 per month \\
1970-1979 &  364 &  10 & 3 per month \\
1980-1989 &  361 &  10 & 3 per month \\
1990-1999 & 1,125 &  10 & 1 per 3 days \\
2000-2006 &  962 &   7 & 1 per 3 days \\
2007-2011 & est. &   5 & 1 per 3 days\tablenotemark{a} \\
2012-2163 & est. & 152 & 1 per day\tablenotemark{b} \\
\enddata
\tablenotetext{a}{Continue 2000-2006 cadence.}
\tablenotetext{b}{Continue 2000-2006 cadence, but assume that LSST and Pan-STARRS come online in 2012 
and each produces one observation every three days for a total observation cadence of one observation per day.}
\end{deluxetable}

In the analysis, we use observation arc lengths running up to 250 years that are evaluated at 50 
year intervals. Since the orbital period of these objects is approximately 250 years, our synthetic 
observations span a complete orbit. The total number of synthetic observations occurring over the 
observation arc lengths are shown in Table \ref{arclength}.

\placetable{arclength}
%\clearpage
\begin{deluxetable}{lrr}
\tabletypesize{\scriptsize}
\tablecaption{\label{arclength}Total number of synthetic observations used in analysis, for each arc segment 
evaluated. The number of observations and their frequency is approximately that actually existing for 
Pluto, with reasonable extrapolations into the future.}
\tablewidth{0pt}
\tablehead{ \colhead{Arc Segments} & \colhead{No. Observations} & \colhead{Cum. Observations} }
\startdata
1914-1963 &    952 &    952 \\
1964-2013 &  4,212 &  5,164 \\
2014-2063 & 18,262 & 23,426 \\
2064-2113 & 18,262 & 41,688 \\
2114-2163 & 18,262 & 59,950 \\
\enddata
\end{deluxetable}

The general approach is to use OrbFit with a set of elements defined as described in Table 
\ref{elements} to generate ephemerides separately in both the perturbed (that is, with the Pioneer 
perturbation present),  and unperturbed (without the Pioneer effect) cases. The predicted positions 
of the test body on the sky represent ``perfect'' observations with no uncertainties in either 
observations or elements. In each case, we can take these predicted ephemeris positions and add 
Gaussian observational error. These randomly altered positions then represent the results of 
synthetic astrometric observations. The random observational error applied is equal to 0.3 seconds 
of arc, representing the results of good quality CCD astrometry reduced against modern star catalogs. 
This value of error is assumed to be isotropic on the sky; thus, we generate and apply synthetic 
errors in right ascension and declination equal to this value divided by the square root of two 
with appropriate adjustment to the Right Ascension for the cosine of the declination. The resulting 
different sets of synthetic observations and alternative gravity models can then be used with OrbFit 
to determine elements and errors associated with the elements as a function of the eccentricity 
and the observation arc length. It is noted that this is the procedure used to investigate the motion 
of minor planets in \citet{2006ApJ...642..606P} and \citet{2007ApJ...666.1296W}. 

We recognize that this assumed error represents a level of accuracy much better than that found in 
observations taken until relatively recently. Indeed, there are many potential sources of position 
error that might be modeled, some of a known nature and some unknown. The data were taken at 
different observatories and different techniques may have been used to reduce the observations, 
leading to systematic errors in the derived positions. The positions have been reduced with various 
catalogs and transformed, sometimes repeatedly, to different reference frames. Even for modern 
observations, there may be errors in the adopted precession values for the J2000 epoch. Additional 
error sources include errors due to uncertainties of the initial conditions of the other planets 
(e.g., ephemeris errors), errors due to corrections to earth rotation, and unmodeled instrumental 
corrections.

Since the primary impact of increasing the a priori positional error is to de-emphasize those 
observations with larger errors, we chose to equally weight all our synthetic observations. Thus, 
we assume much more optimistic error estimates for early observations than are representative in 
the actual observation archive and results in equal weighting for all the synthetic observations. 
This should result in our calculations providing an optimistic estimate of the detectability of 
the Pioneer perturbation.

An additional reason for electing to model a uniform observational accuracy across the span of 
the synthetic observations in order to minimize variation in our results due to another source of 
observational noise. For parts of the synthetic observational arc prior to the advent of modern 
catalogs and CCDs, this error will be considerably less than that found in the actual observational 
archive. Interestingly, the OrbFit program is capable of varying a priori observational error 
across time or across different observatories, but we did not make use of this capability in this 
paper.

Given these assumptions, we have four cases, which we illustrate in Table \ref{cases}. The first 
pair of cases (the ``Gravity with PE'' column in Table \ref{cases}), and the focus of our analysis, 
is for observations that are generated with a Pioneer-effect perturbation present meaning that the 
Pioneer effect exists in Nature and determines an object's motion. These same observations are 
analyzed with two gravity models. The first model is that of Newtonian gravity with the addition of 
a constant radial acceleration with a value commensurate with that associated with the Pioneer 
effect; the second is standard Newtonian gravity. The comparison of the orbital solutions for these 
two cases correspond to a situation in which the universe is one where the Pioneer effect actually 
exists and we analyze it to determine which of our two gravity models is correct. This is the 
situation portrayed in the first column of Table \ref{cases}. The second pair of cases are comparable 
except that they are associated with a universe where the Pioneer effect does not exist. Again, we 
compare two theories of gravity to see if we can observationally distinguish between the two. This 
situation is portrayed in the second column of Table \ref{cases}. This paper investigates the first 
case only; as shown in \citet{2006ApJ...642..606P}, the second case is nearly symmetric with the 
first, and produces similar conclusions.

\placetable{cases}
%\clearpage
\begin{deluxetable}{lcc}
\tabletypesize{\scriptsize}
\tablecolumns{3}
\tablewidth{0pt}
\tablecaption{\label{cases}Four cases combining gravity models and forces determining motion.}
\tablehead{\colhead{} & \multicolumn{2}{c}{Forces Determining Motion\tablenotemark{a}} \\ \cline{2-3} \\
           \colhead{Gravity Model\tablenotemark{b}} & \colhead{Gravity with PE\tablenotemark{c}} & 
           \colhead{Only Gravity\tablenotemark{d}}}
\startdata
Gravity with PE\tablenotemark{c} & ``Matched'' & ``Mismatched'' \\
Only gravity\tablenotemark{d} & ``Mismatched'' & ``Matched'' \\
\enddata
\tablenotetext{a}{By this we mean that the motion of the orbiting objects are determined by the forces 
indicated. The first column indicates that there is really a Pioneer effect perturbation superimposed 
on Newtonian gravity; the second column indicates that there is not an additional perturbation.}
\tablenotetext{b}{``Gravity Model'' refers to the assumed force law to which the orbital observations 
are fit. The first row corresponds to the case where the assumed force law is Newtonian gravity with 
an additional Pioneer-like constant acceleration; the second row refers to a force law purely Newtonian 
in character.}
\tablenotetext{c}{``Gravity with PE'' indicates that there is a Pioneer effect perturbation.}
\tablenotetext{d}{``Only gravity'' indicates that there is no Pioneer effect perturbation.}
\end{deluxetable}

\section{\label{results}RESULTS}

\subsection{Prediction of sky position from orbital elements} 

\subsubsection{Projected orbits} 

As outlined above, our plan is to take ephemerides created with various known orbital elements and 
predict positions on the sky when the motion is governed by Newtonian gravity with-- and without a 
Pioneer-like perturbing acceleration. This process involves integrating the equations of motion of 
a body subjected to two different force laws. 

\begin{figure*}
\plottwo{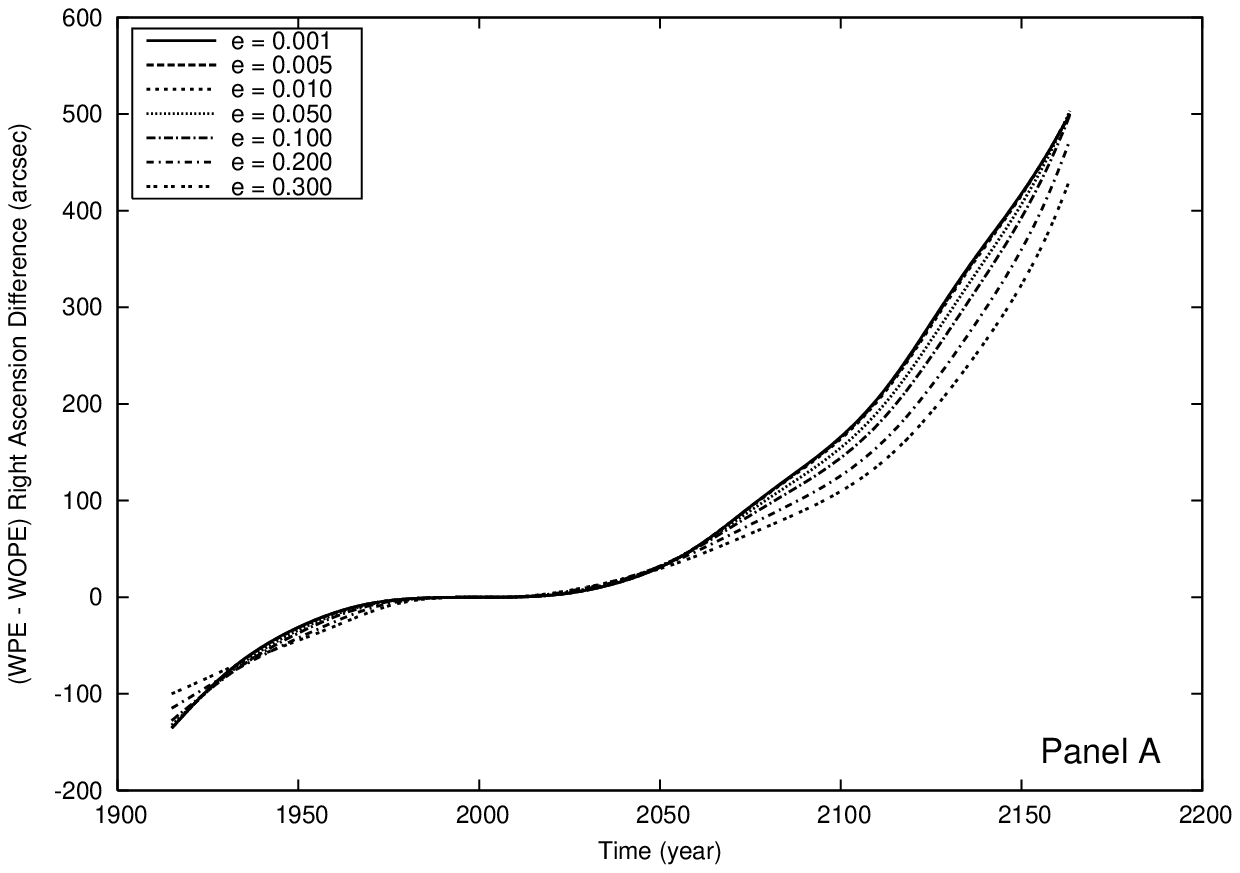}{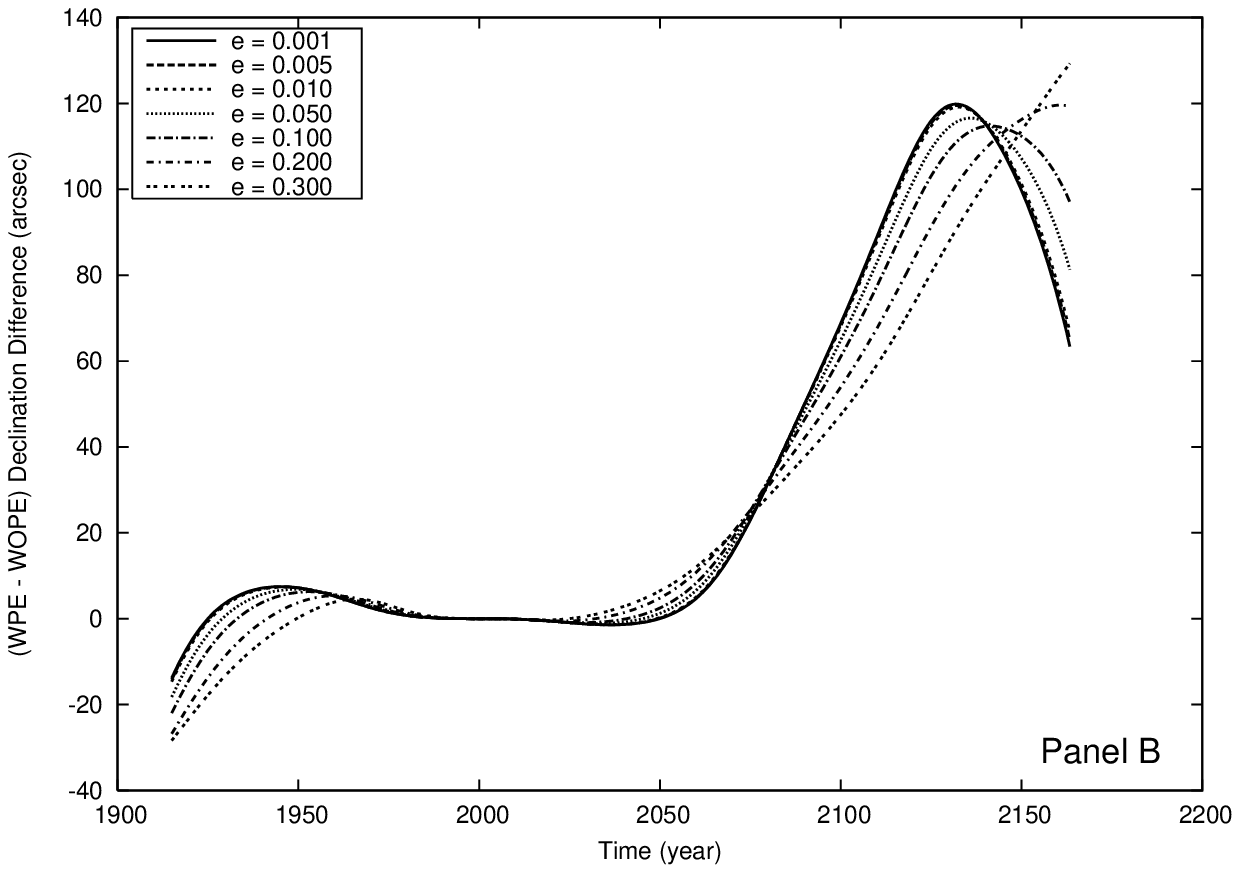}%
\caption{\label{fig1}Angular position differences when orbits are extrapolated with ``known'' 
elements with-- and without a Pioneer effect perturbation. The orbits have elements given in 
Table \ref{elements}, except for the eccentricity which is parametrically varied. Panel A (left) 
shows the right ascension difference between the perturbed and unperturbed cases. Panel B (right) 
similarly shows the declination difference. In both cases, the difference is zero at 2000.0 because 
that is the epoch of the elements used in the calculation.}
\end{figure*}

The results of this calculation provide sky positions as a function of time. Since the angular 
differences between the predicted positions are small, we separately consider the difference in 
right ascension and declination resulting from the two situations. Fig. \ref{fig1} shows the 
resulting position difference. Since the epoch of the elements is JD 2,451,545.0, the angular 
differences are zero at that date and diverge as one moves forward and backwards in time from 
the element epoch. It is worth emphasizing that the two ephemerides being compared are generated 
with two separate gravity models with identical and exactly specified elements. 

Figure \ref{fig1} can be compared with Figs. 9 and 10 in \citet{2006NewA...11..600I}. They drew 
the conclusion that the divergence in position, since not observed over the past 90 years, indicates 
the Pioneer effect does not exist. However, such a conclusion is based on the idea that the elements 
(i.e., the initial conditions) are known precisely. The reality is that these position predictions 
are those related to the elements derived from the assumption of Newtonian gravity coupled with 
observed positions. If there is a ``mismatch'' between the gravity model used to derive orbital 
elements and the reality that determines the actual motion of the orbital bodies, these will not 
be the correct elements. Even if the elements are derived from a ``matching'' reality and gravity 
model, it is remarkable how rapidly orbital predictions can degrade outside the observation arc, 
especially when that arc is short relative to the orbital period. These aspects will be discussed 
in detail in the next subsection.

\subsubsection{Fitted orbits} 

If we take the ephemeris positions associated with one ``reality'' (e.g., the one containing the 
Pioneer perturbation---see Table \ref{cases}) consonant with the observation cadence shown in Table 
\ref{obsfreqs}, and add an isotropic, normally distributed random position error of 0.3 arc seconds 
to each of the positions, the result is a set of synthetic positions that are illustrative and 
representative of those that might have been obtained as observations of the bodies. This set of 
observations can be divided into five different arcs covering 250 years (approximately one orbital 
period); thus, each arc covers an additional 50 years from the start of the observations and 
represents an incremental one-fifth of an orbital revolution.

Each synthetic observation arc can be analyzed when a Pioneer-like perturbation is included and when 
it is not, corresponding to the two rows in the first column of Table \ref{cases}. The observation 
fitting process serves to determine the elements of the orbit and these are the elements that can 
then be used to predict sky position. How do the observed positions differ when this procedure is 
followed?

\begin{figure*}
\plottwo{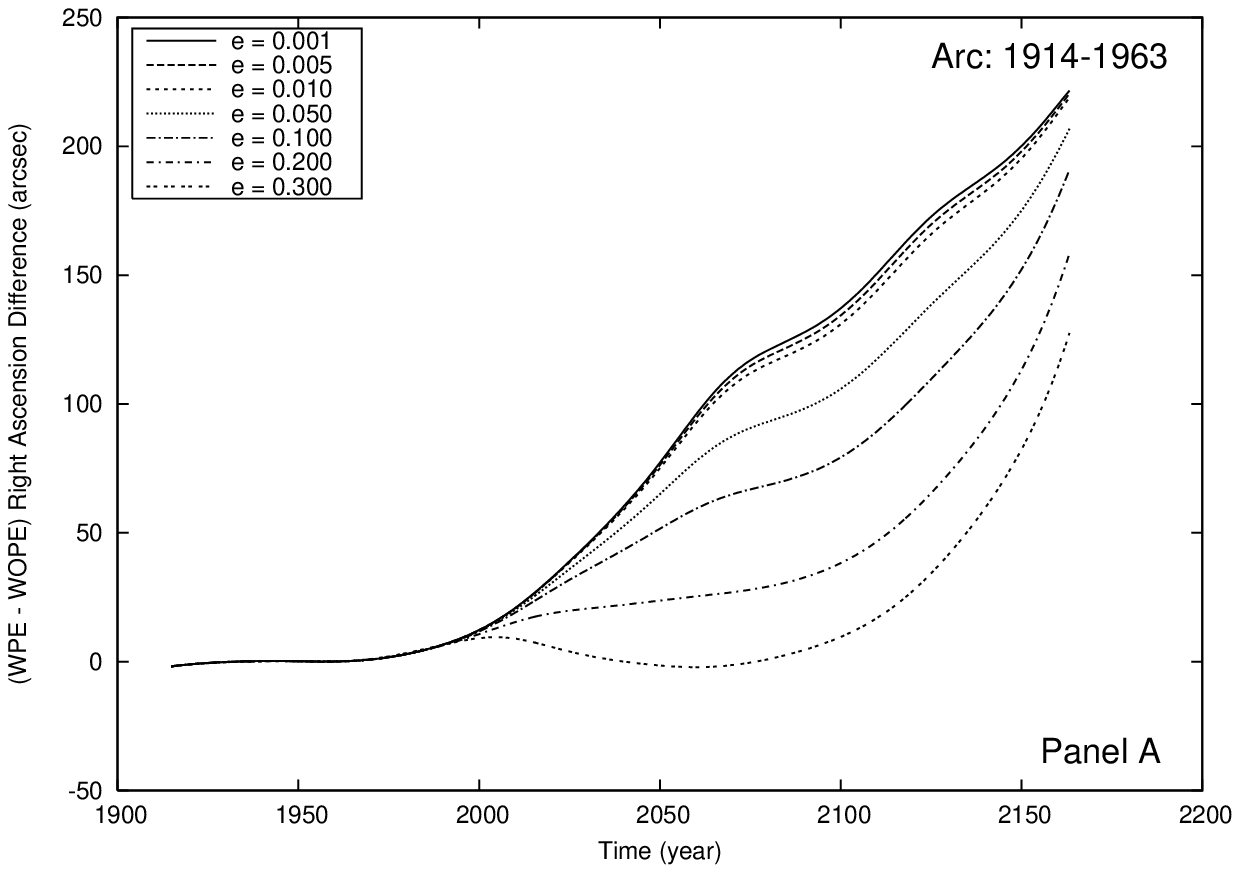}{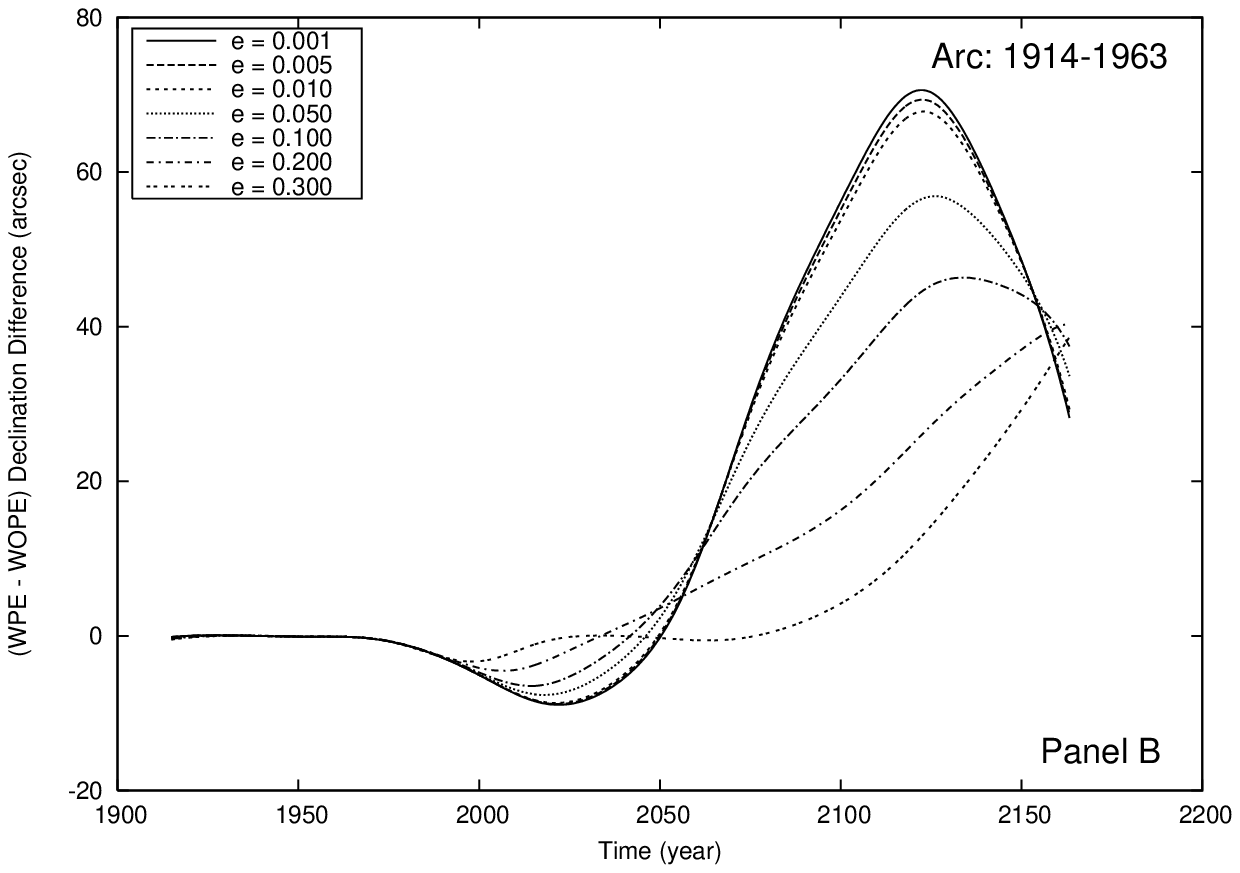}
\caption{\label{fig2}Angular position difference when orbits are extrapolated with elements 
determined from synthetic observations generated with a Pioneer effect perturbation. The elements 
are those given in Table \ref{elements} except for eccentricity which is varied parametrically. 
The observation arc is of 50 years duration, through 1963. Panel A shows the difference in right 
ascension between determining the orbital position with a gravity model including the perturbation 
and one not including the Pioneer effect. Panel B shows the similar declination difference. For 
the entire length of the observation arc there are only very small differences between the models.}
\end{figure*}

\begin{figure*}
\plottwo{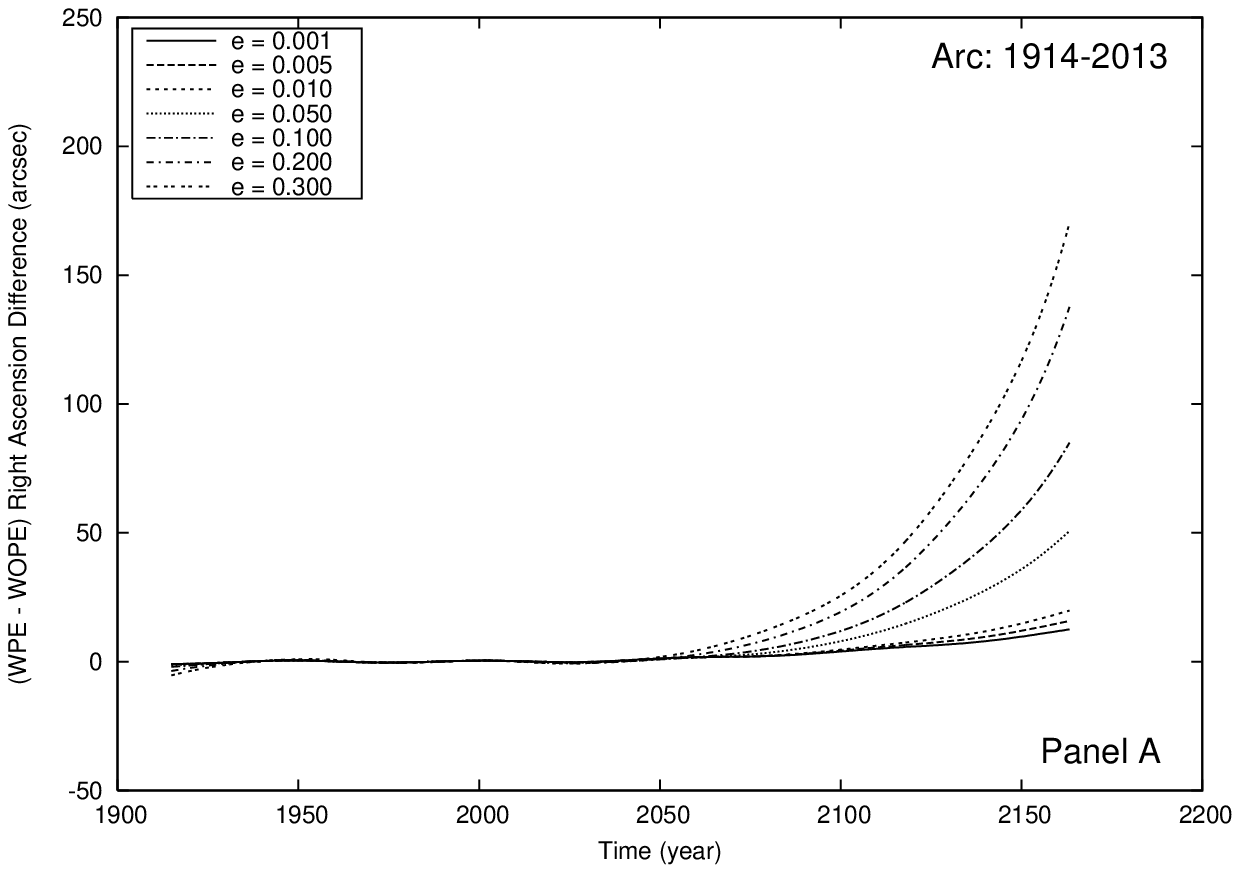}{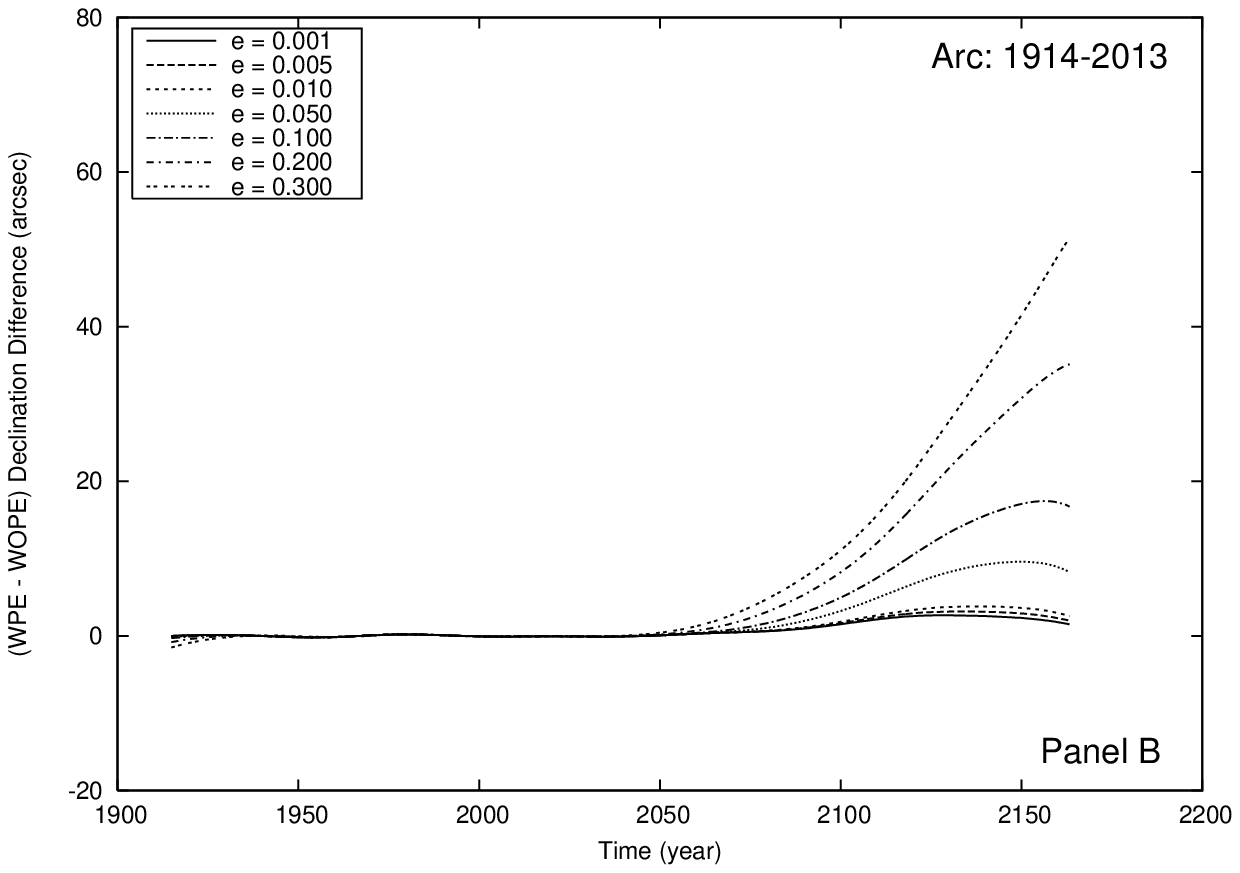}%
\caption{\label{fig3}Angular position difference when orbits are extrapolated with elements 
determined from synthetic observations generated with a Pioneer effect perturbation. The elements 
are those given in Table \ref{elements} except for eccentricity which is varied parametrically. 
The observation arc is of 100 years duration, through 2013. Panel A shows the difference in right 
ascension between determining the orbital position with a gravity model including the perturbation 
and one not including the Pioneer effect. Panel B shows the similar declination difference. Only 
very small differences in predicted position can be detected over the length of the observation 
arc and are discussed in detail in the text.}
\end{figure*}

\begin{figure*}
\plottwo{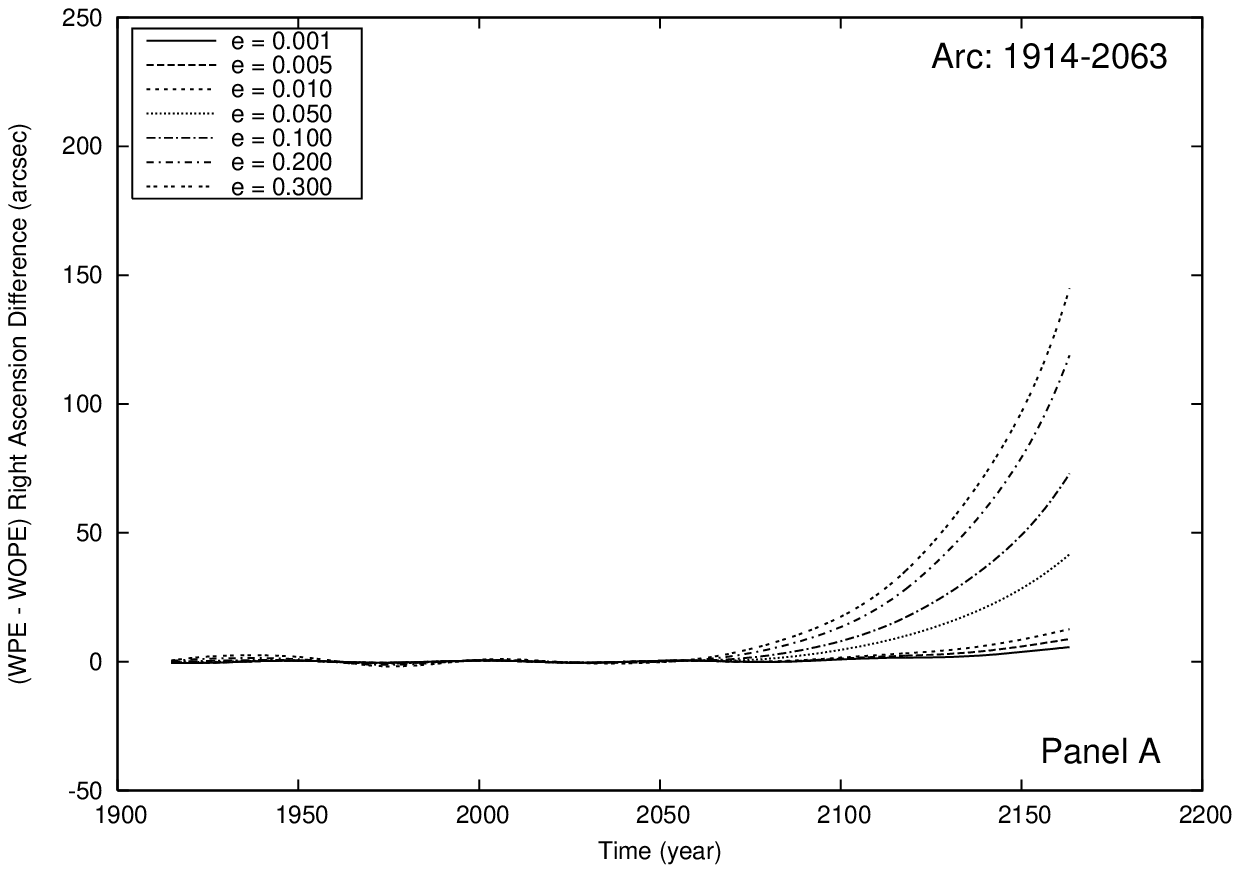}{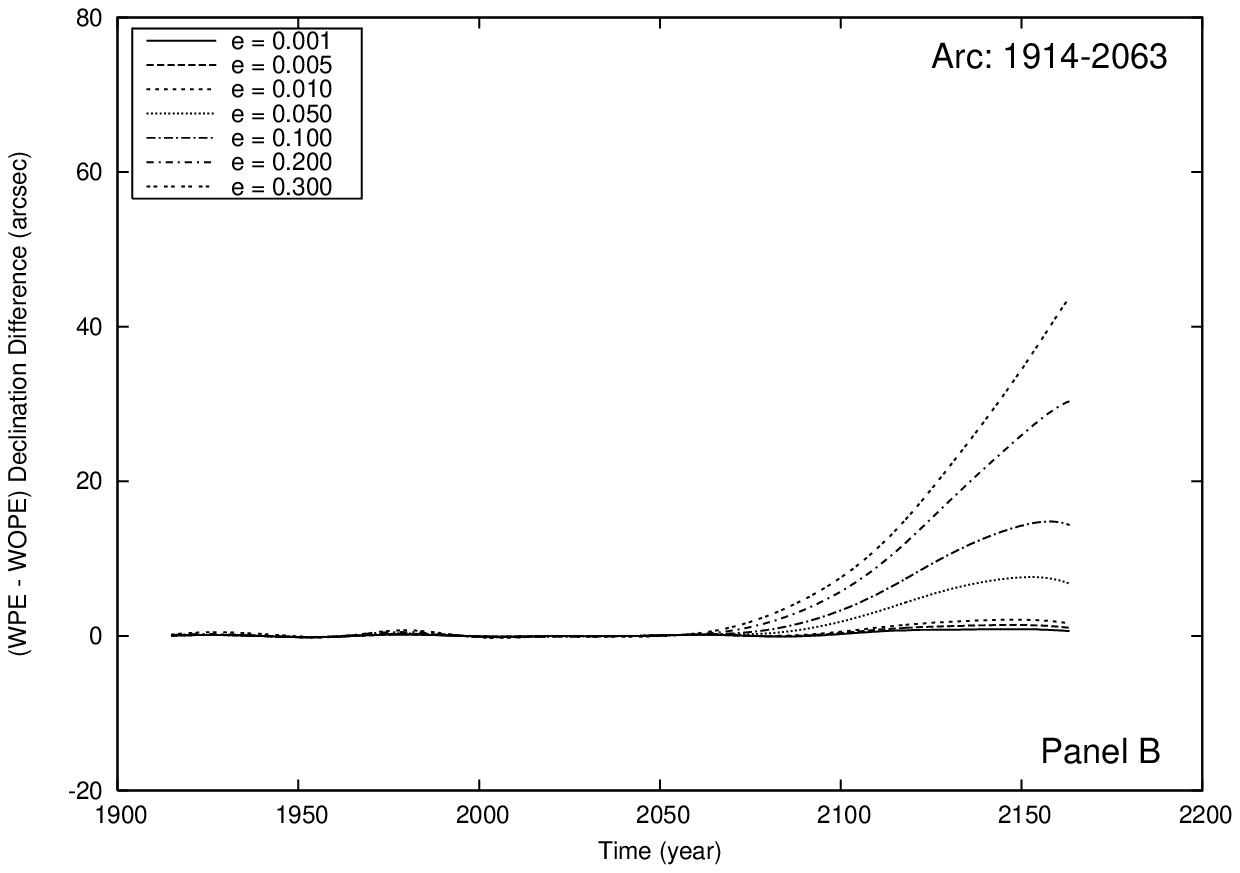}%
\caption{\label{fig4}Angular position difference when orbits are extrapolated with elements 
determined from synthetic observations generated with a Pioneer effect perturbation. The elements 
are those given in Table \ref{elements} except for eccentricity which is varied parametrically. 
The observation arc is of 150 years duration, through 2063. Panel A shows the difference in right 
ascension between determining the orbital position with a gravity model including the perturbation 
and one not including the Pioneer effect. Panel B shows the similar  declination difference. Very 
slight position differences between the two gravity models can be discerned within the observation 
arc.}
\end{figure*}

\begin{figure*}
\plottwo{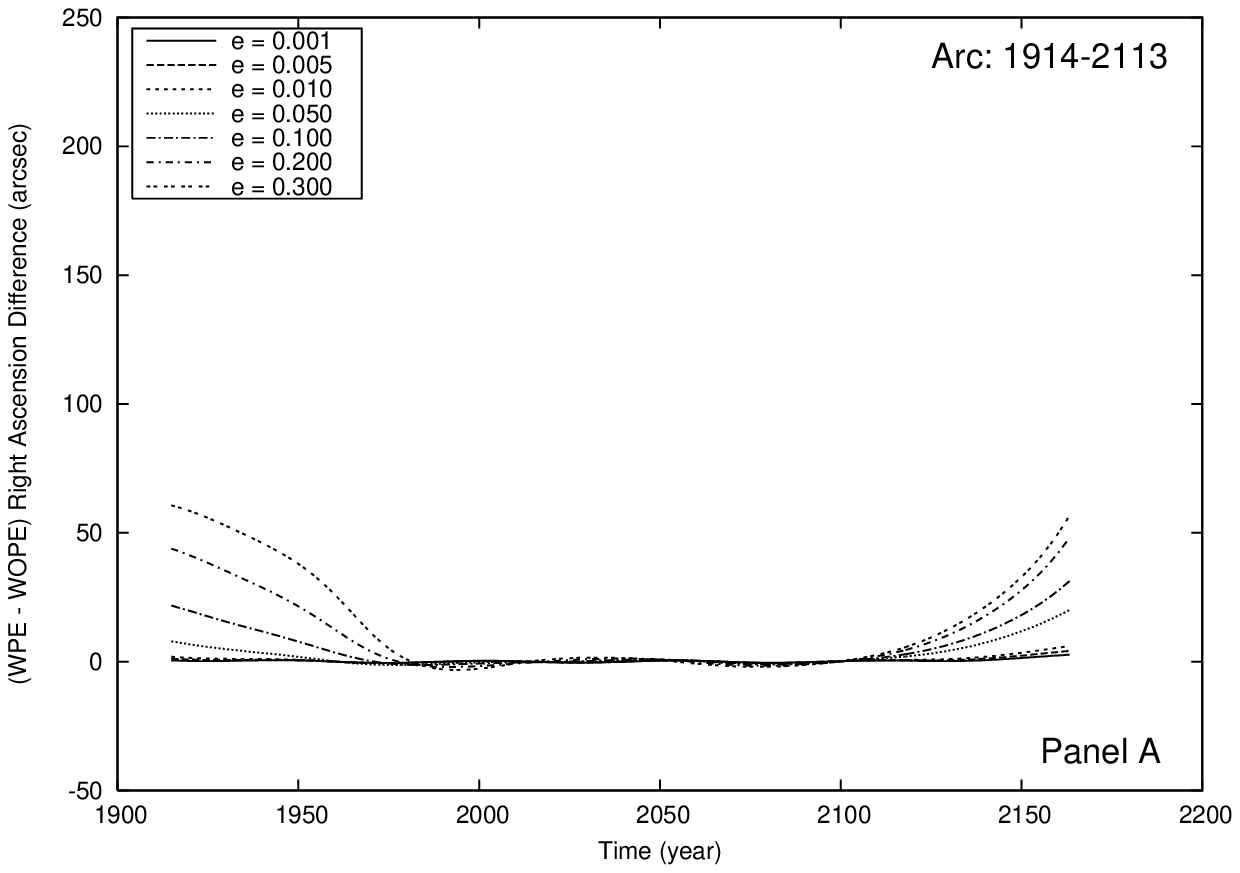}{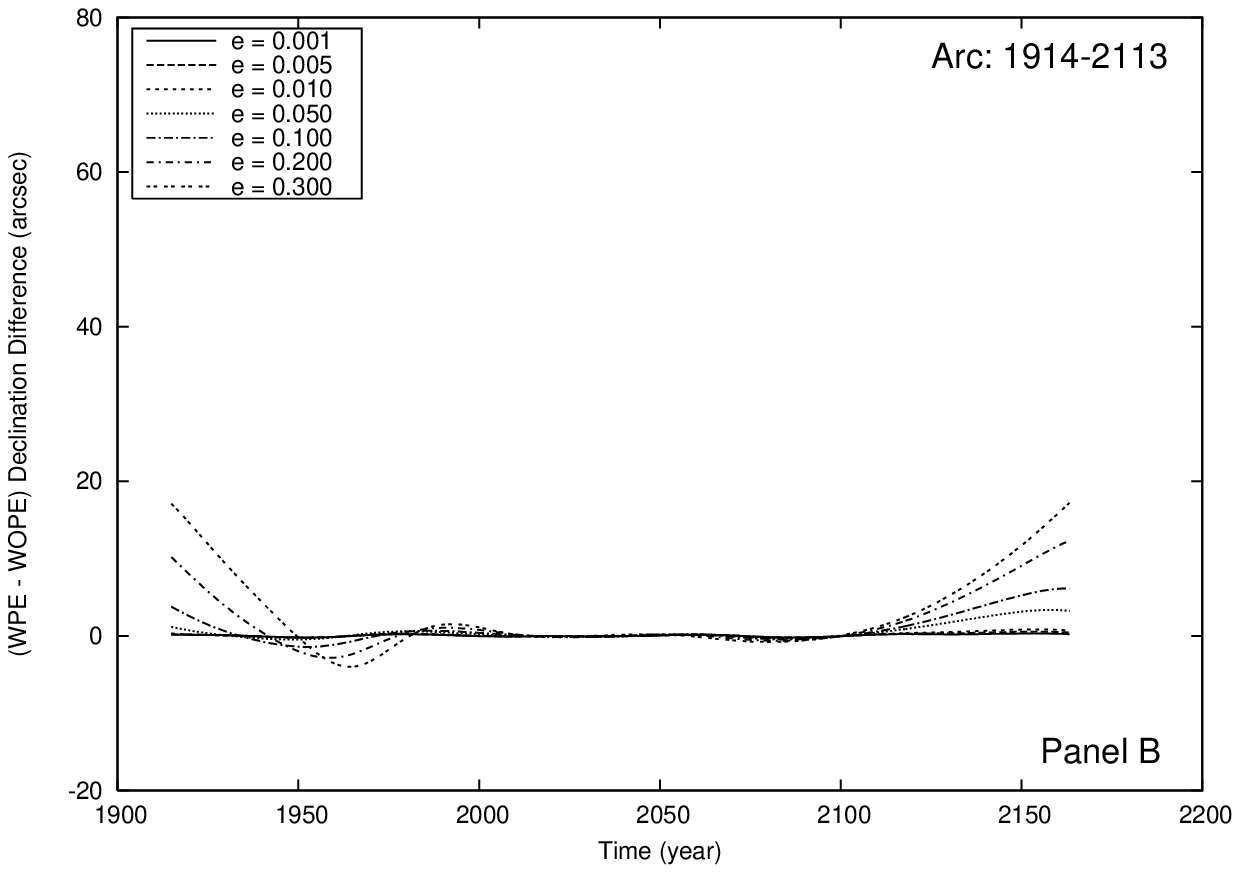}%
\caption{\label{fig5}Angular position difference when orbits are extrapolated with elements 
determined from synthetic observations generated with a Pioneer effect perturbation. The elements 
are those given in Table \ref{elements} except for eccentricity which is varied parametrically. The 
observation arc is of 200 years duration, through 2113. Panel A shows the difference in right 
ascension between determining the orbital position with a gravity model including the perturbation 
and one not including the Pioneer effect. Panel B shows the similar declination difference. 
Substantial differences between the positions predicted by the two gravity models can now be seen 
inside the observation arc, allowing one to falsify at least one of the gravity models.}
\end{figure*}

\begin{figure*}
\plottwo{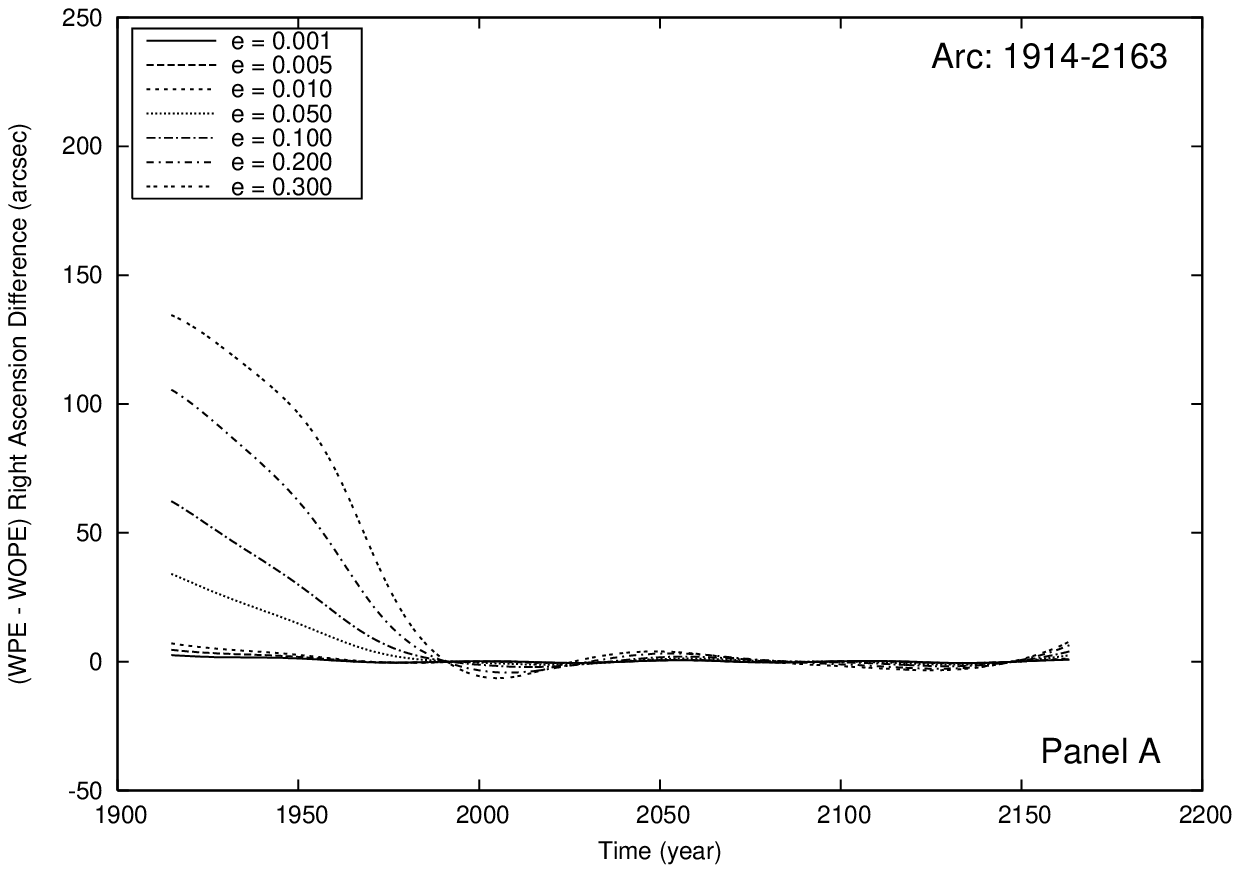}{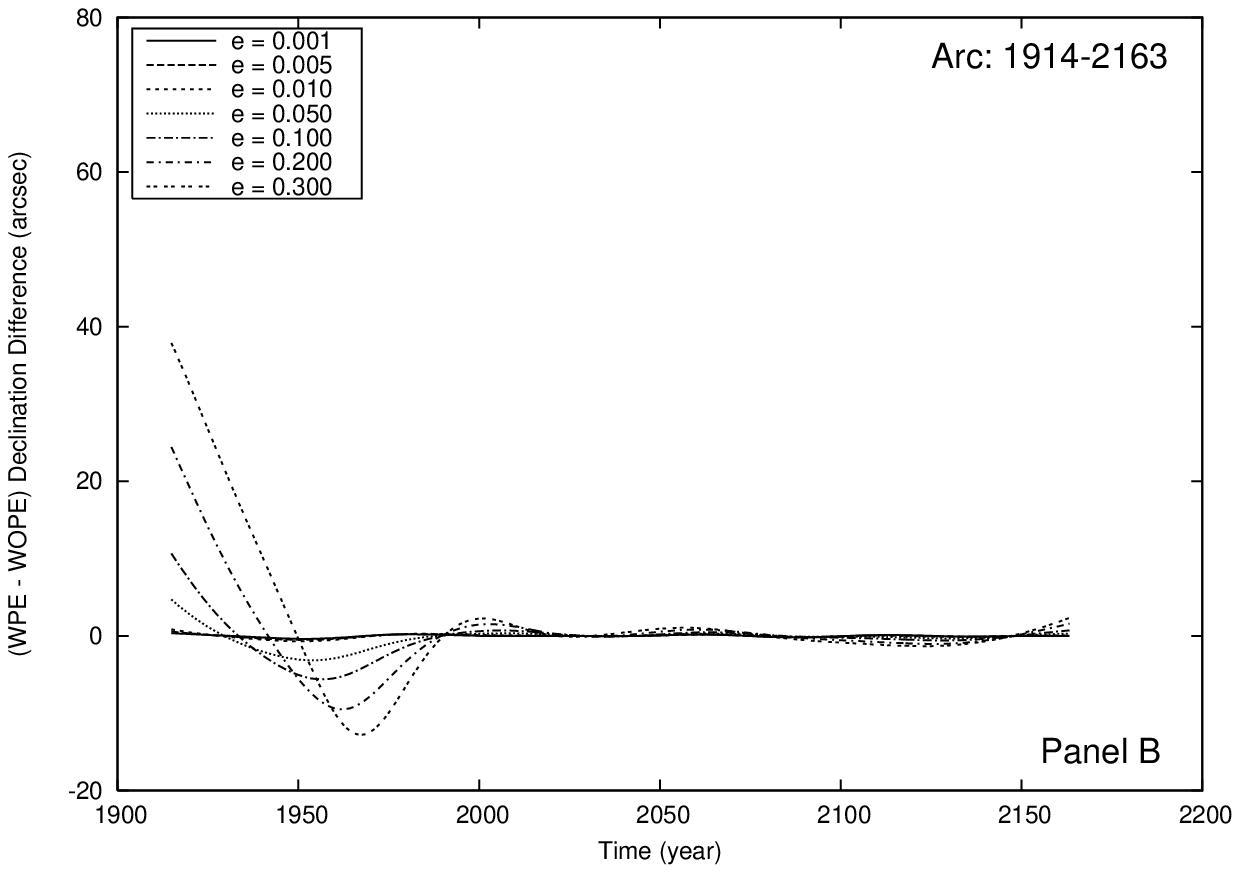}%
\caption{\label{fig6}Angular position difference when orbits are extrapolated with elements 
determined from synthetic observations generated with a Pioneer effect perturbation. The elements 
are those given in Table \ref{elements} except for eccentricity which is varied parametrically. The 
observation arc is of 250 years duration, through 2163. Panel A shows the difference in right 
ascension between determining the orbital position with a gravity model including the perturbation 
and one not including the Pioneer effect. Panel B shows the similar declination difference. The 
trends in fit quality seen in the 200 year arc continue. Large differences in predicted position 
allow at least one gravity model to be falsified.}
\end{figure*}

Figs. \ref{fig2} through \ref{fig6} shows the difference in sky position as a function of time for 
the orbital fits associated with the two gravity models. In all cases, the left hand graph shows the 
right ascension difference while the right hand graph shows the declination difference. Comparing these 
figures with Fig. 1 above and Figs. 9 and 10 in \cite{2006NewA...11..600I} shows that for shorter 
arcs (those ranging up to 100 years long) the difference in predicted motion for the two gravity 
models is quite small for the duration of the observation arc and only begins to substantially 
diverge when the position is being extrapolated to times beyond those for which observations 
exist. For a 150 year arc, small irregularities in the position difference begin to be noticeable 
during the observational arc. As the arcs grow still longer (to 200 years and beyond), more 
substantial irregularities are seen. Also, it is generally true that positional differences 
are greater the greater the eccentricity, all other things being equal.

Recall what the differences plotted in Figs. \ref{fig2} through \ref{fig6} represent. Synthetic 
observations for a universe with a Pioneer effect were generated and orbital elements were determined 
by fitting the same observations to two alternative gravity models. One of these models is correctly 
``matched'' to the observations; that is, the gravity model used for orbit fitting is the same one 
used to generate the observations. The other model does not correspond to the one used to generate 
the observations. Thus, we would a priori expect that the former would fit the data better than the 
latter.

What we see, however, is surprising. For observation arcs as long as a century, which at best is the 
situation obtaining for Pluto at the current time, there are only relatively small differences in the 
positions predicted by the correct gravity model and those predicted by the mismatched model. In all 
cases, however, the predicted positions outside the range of observations rapidly diverge from one 
another.

This behavior is a manifestation of two inter-related factors. First, the orbital fitting problem 
is inherently nonlinear and is normally solved in the linear approximation. Even if not 
mathematically chaotic, the system of equations is certainly sensitively dependent upon initial 
conditions; small changes in elements can result in large changes in predicted position outside the 
range of observations. Secondly, this sensitivity is exacerbated by the problem of a short observation 
arc. The length of the entire observational archive for Pluto is no more than about one-third of its 
orbital period. Even in as simple a case as linear least squares curve fitting, a limited amount of 
independent variable data (corresponding to a short observation arc) will lead to relatively large 
errors in the fitting parameters (corresponding to the orbital elements). Together, these factors 
conspire to potentially generate large position errors outside the observation arc, while increasing 
the length of the observation arc can markedly reduce error over the whole of the arc and even beyond 
it.

Some of these issues can sometimes be alleviated by a change in variables. Use of other than 
Keplerian orbital elements (e.g., equinoctial elements) can allow one to avoid some problems with 
determining an orbital solution, in particular in cases with small eccentricities or small 
inclinations. Similarly, a linear combination of Keplerian orbital elements can be used to improve 
the accuracy of some elements (but see the footnotes on pages \pageref{fn1} and \pageref{fn2}). 
However, those types of variable change do nothing for the issues associated with the ill-conditioned 
nature of the linear matrix and a short observational arc. 

In any case we see that we must fit observations to particular gravity models and adjust orbital 
elements before predicted positions on the sky can be compared. Noting that deriving orbital 
elements and predicting an orbit beyond a short observation arc provides little information about 
the actual motion that will be seen once new observations are made, we conclude that drawing 
conclusions from such extrapolations is totally unwarranted. Thus, we can conclude nothing 
regarding the Pioneer effect and the motion of Pluto as it is currently known, at least subject 
to the limitations of our methodology. 

\begin{figure*}
\plottwo{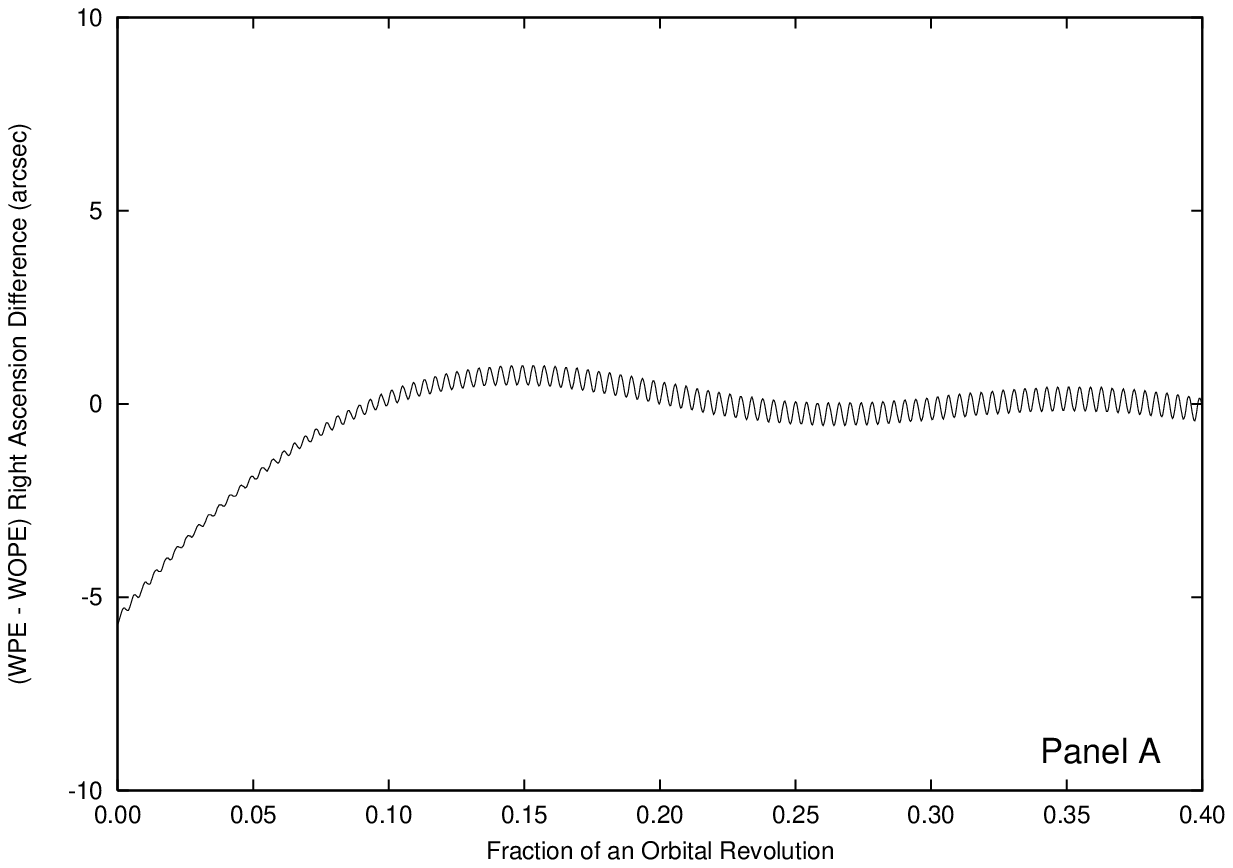}{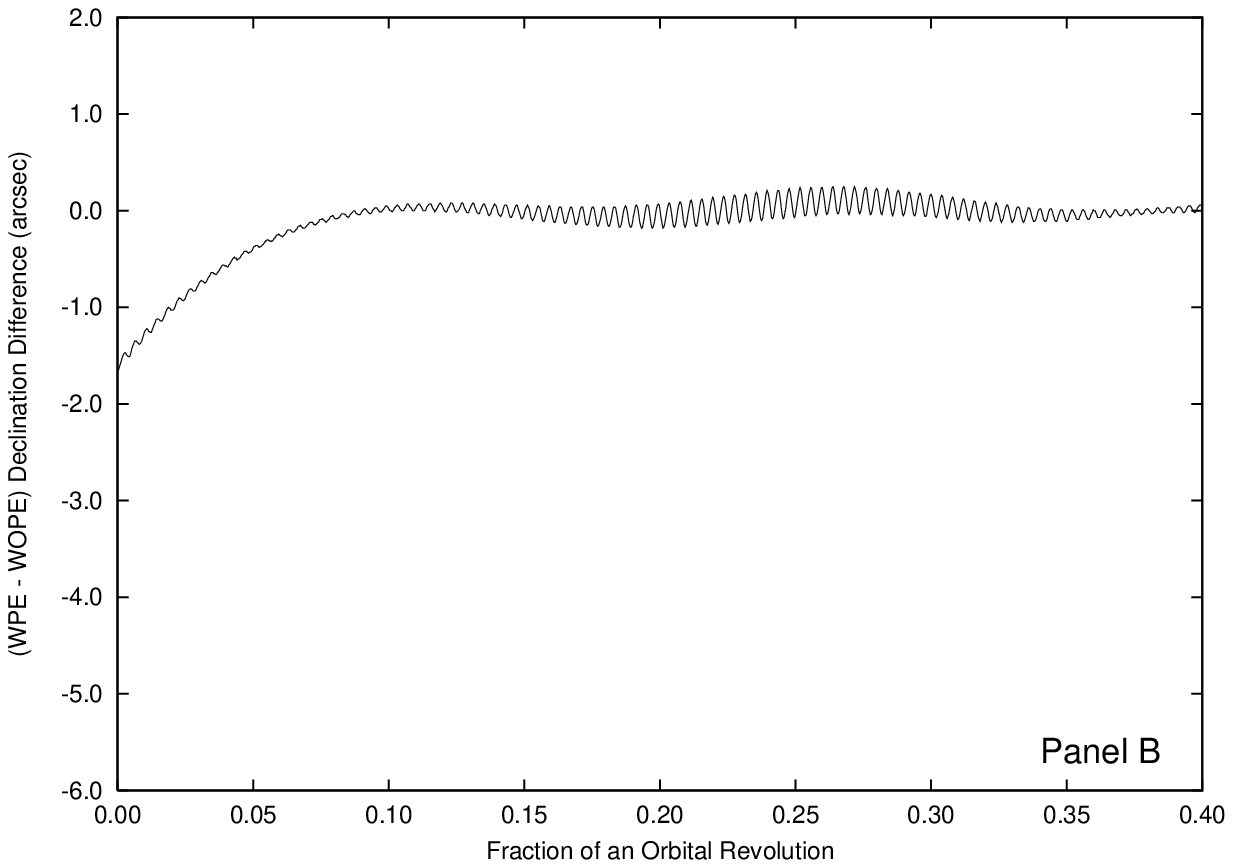}%
\caption{\label{fig12}Angular position differences for Pluto when orbits are predicted with elements 
determined from synthetic observations generated with a Pioneer effect perturbation. The time axis 
is labeled in units of Pluto's orbital period running from the beginning of modern observations up 
to the present. Panel A shows the difference in right ascension between determining the orbital 
position with a gravity model including the perturbation and one not including the Pioneer effect. 
Panel B shows the similar declination difference. One would think that this discrepancy would be 
observable; however, we argue in the text that it is not a distinguishable difference given our 
current knowledge of Pluto's orbit.}
\end{figure*}

Now let us examine more closely the relatively small position differences at the beginning of the 
observation arcs in Figs. \ref{fig2} through \ref{fig6}. Fig. \ref{fig12} is similar to Figs. 
\ref{fig2} through \ref{fig6}, but shows an expanded view of the difference in position of a body 
with an eccentricity of 0.3 (e.g., similar to Pluto) and a 100 year observation arc. At the beginning 
of the arc, the difference in right ascension for the two gravity models is nearly six seconds of 
arc and the difference in declination is about 1.5 arcsec. One would certainly think that this level 
of difference would be observable; however, we argue below that it is not a distinguishable 
difference given our current knowledge of Pluto's orbit.

\begin{figure*}
\plottwo{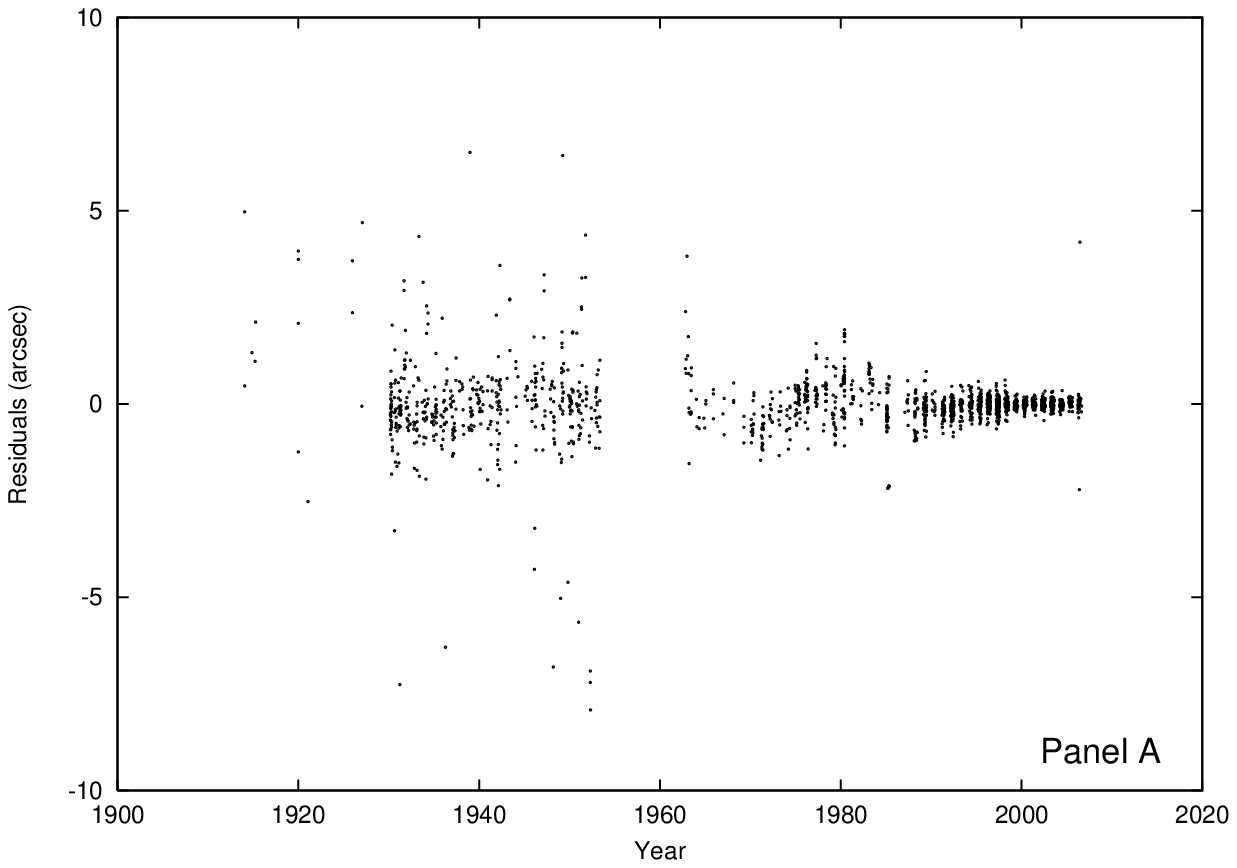}{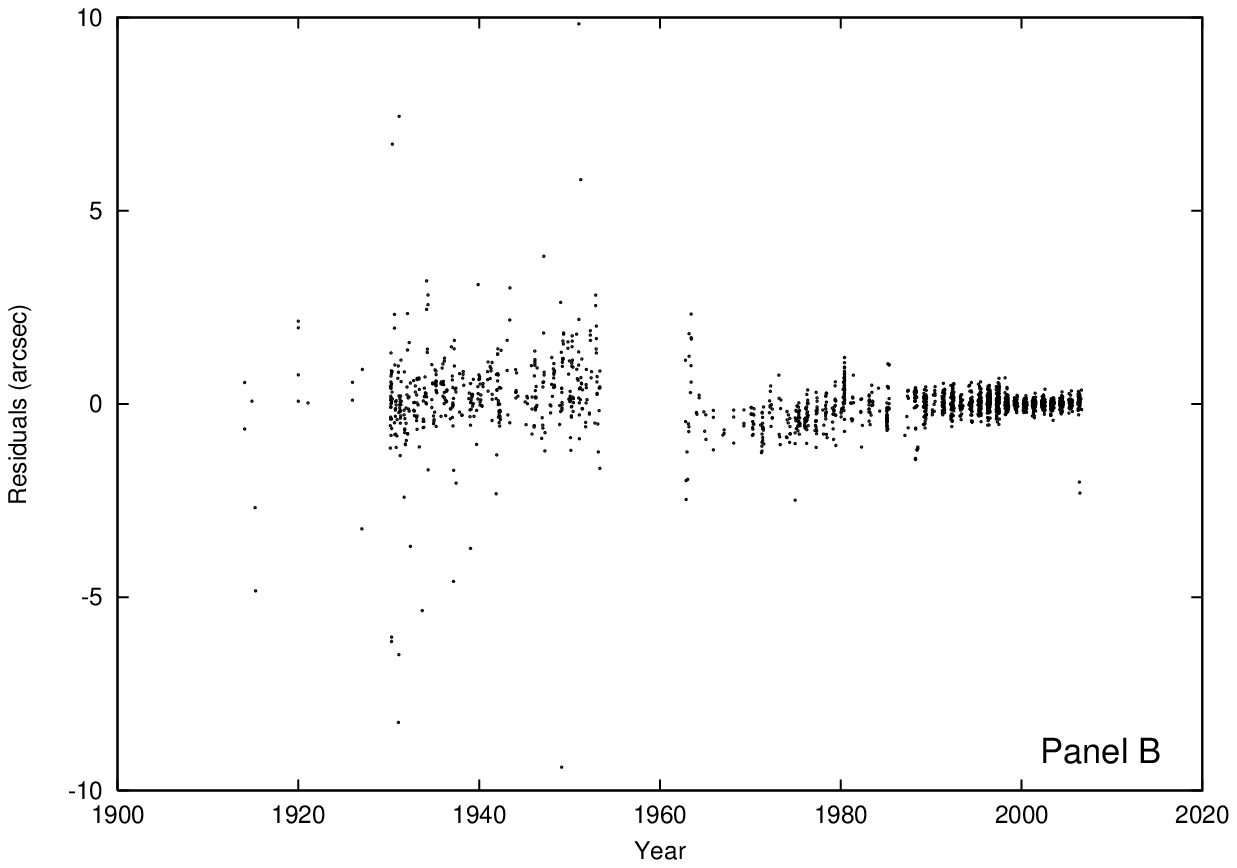}%
\caption{\label{fig13}Observed minus calculated residuals for Pluto with respect to the DE414 
ephemeris. Panel A provides the right ascension residuals in arc seconds while Panel B gives the 
declination residuals. The same scale is used in this figure as in Fig. \ref{fig12}.}
\end{figure*}

Figure \ref{fig13} shows a subset of the residuals in both right ascension and declination of actual 
observations with respect to a recent JPL ephemeris, DE414. The full set of residuals show a number 
of clear outliers and we have removed from the set any residuals of magnitude greater than ten seconds 
of arc. Prior to about 1960 there is clearly a greater spread in the residuals than at later times. 
Additionally, prediscovery images taken prior to 1930 and dating to as early as 1914 have been found 
in the archive. These observations seem to contain serious irregularities manifested by large and 
systematically biased residuals. Supporting this assertion, \citet{1994Icar..108..174G} find that 
systematic differences between the observed and calculated orbit are a continuing issue. Additionally, 
they find that only about half of Pluto's motion since its discovery has been observed in a systematic 
and organized manner. They also comment that it is impossible to extrapolate Pluto's position more than 
a few years into the future. This is indicative of a poorly characterized orbit. 

The residuals of the synthetic observations with respect to a ``mismatched'' gravity model are similar 
in character and magnitude to the differences between predictions for different gravity models that are 
shown in Fig. 7. As observed previously, there seems to be a significant difference between the two sets 
of predictions. A standard comparison of these models \citep{2003drea.book.....B} would calculate and 
compare $\chi^{2}$ values for the two models. If the resulting value is about unity, the fit is good. 

The overall issues associated with model goodness-of-fit are complex and will be discussed later 
in Section \ref{quality}; however, anticipating this complexity let us look more deeply at the data. 
If we compare our ``mismatched'' orbit fit with actual observational data and their associated fit, 
we will see that there are comparable systematic errors in both the synthetic and the real datasets. 

\subsubsection{Simulated post-fit residuals with respect to the DE414 ephemeris}

\begin{figure*}
\plottwo{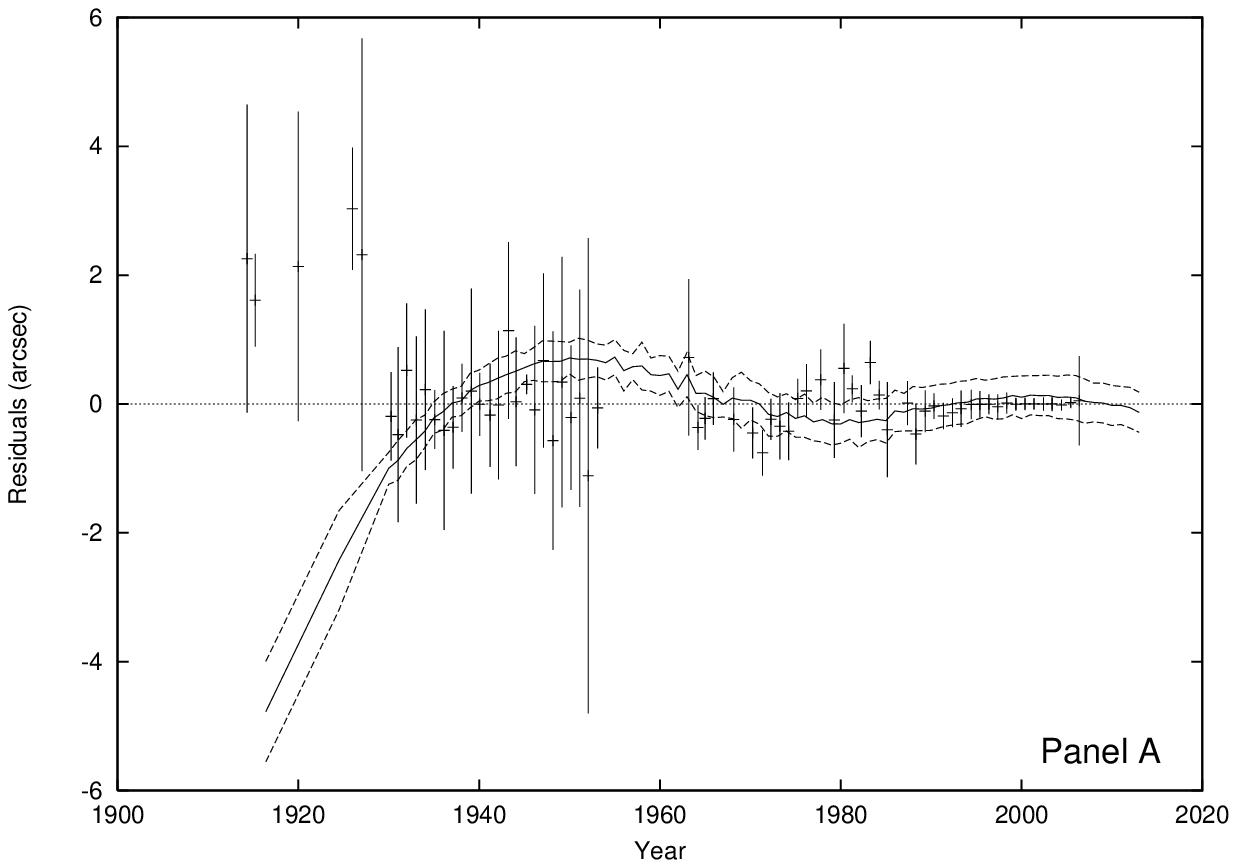}{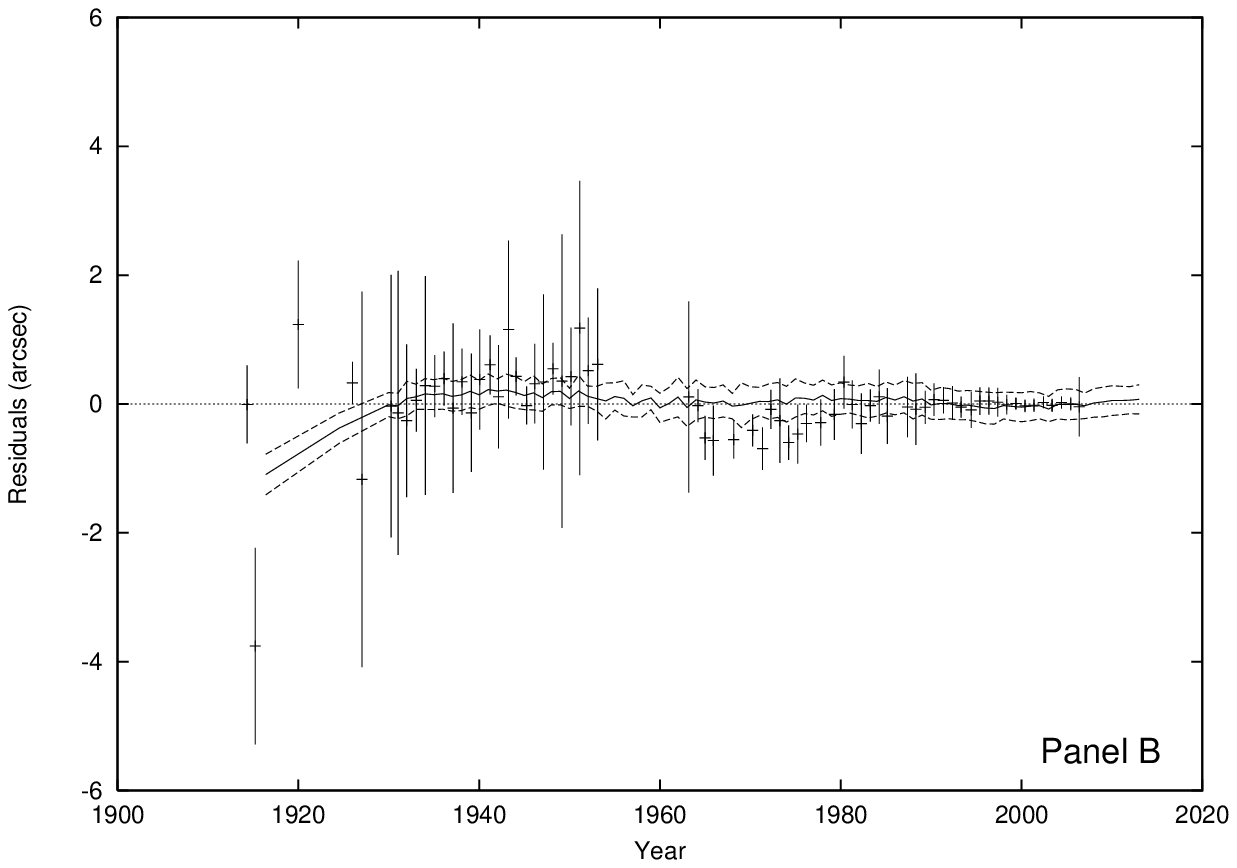}%
\caption{\label{fig14}Normal points for the DE414 residuals and the residuals for the synthetic 
observations relative to their ephemeris. Panel A pertains to the right ascension residuals while 
Panel B applies to the declination residuals. The ``high-low-close'' portion of the graph shows 
normal points for the DE414 residuals. The lines show normal points and a one standard deviation 
variation about the mean for the residuals of the synthetic observations relative to their 
ephemeris.}
\end{figure*}

Systematic trends in residuals can be visually discerned more readily by examination of normal 
points rather than individual data points. In a normal point plot, means and standard deviations 
of the residuals are collected and plotted for each planetary opposition. Fig. \ref{fig14} shows 
the normal points for the DE414 residuals in the ``high-low-close'' portion of the plot, where the 
``closing'' value indicated by the small horizontal mark is the average residual for each opposition 
and the length of the vertical line indicates a one standard deviation variation above and below 
the mean residuals. The comments above about the systematic bias of the prediscovery observations 
and the larger dispersion of residuals before 1960 are borne out in the normal plot. 

Now, let us compare this data with normal points for the synthetic observations of Pluto. In 
developing the observation cadence for the synthetic observations, we did not consider the 
relative positions of the Earth, the Sun, and Pluto and thus did not concentrate synthetic 
observations around particular yearly oppositions. Since our primary purpose was to illustrate 
broader issues with the data, this was done to avoid introducing additional noise through a 
varying observational cadence. Thus, for the post-discovery residuals associated with our 
synthetic observations we found the average and standard deviations for each calendar year. 
For prediscovery images, our assumed observation cadence of one observation per year clearly 
made the dispersion calculation impossible. In these cases, we accumulated data by decade; 
thus, we calculated the mean and standard deviation for the 1914-1919 period and the 1920-1929 
period and plotted the resulting values accordingly.

The normal points associated with the synthetic observations are also shown in Fig. \ref{fig14}. 
The average is shown by the solid line and the plus and minus one standard deviation limits are 
shown as dotted lines. Several comments are in order about these graphs. First, the width of the 
standard deviation spread on the synthetic observations is different from that of the DE414 
residuals. This is due to our assumption of a constant 0.3 arc second astrometric accuracy of 
the synthetic observations. The actual observations that contribute to the dispersion of the 
DE414 residuals have varying accuracies, ranging from large values for the early observations to 
small values for current CCD observations. As pointed out earlier, the level of astrometric 
accuracy used in the synthetic observations is much better than that found in the  actual 
observation archive until relatively recently. We chose to model a uniform observational 
accuracy across the span of our synthetic observation arc in order to make an optimistic estimate 
of the detectability of the Pioneer perturbation as well as to minimize variation in our results 
due to another source of noise (e.g., the observation error). Thus, rather than the uniform 
difference between the upper and lower limits, the normal points of the actual DE414 residuals 
vary in width. If we had modeled a varying error, such a variation in width would be expected 
for the synthetic observations as well.

Another point to be made about Fig. \ref{fig14} is that while the prediscovery residuals for the 
synthetic observations and those with respect to DE414 are of opposite signs, their magnitudes 
are of the same order. In the former case, the large residuals are due to fitting a ``mismatched'' 
gravity model consisting of only nominal gravity to observations that result from gravity plus an 
anomalous Pioneer acceleration; in the latter case, unknown observational errors have resulted in 
large and biased residuals. What are we to make of the difference in sign? We would argue that 
since there are errors of an unknown nature present in these early observations, we have absolutely 
no basis is even assuming the sign of the residual is indicative of reality. Although the 
prediscovery residuals have characteristics that indicate that they might represent a constant 
offset in right ascension, an effect that can be caused by many errors in the analysis of the 
observations, the limits on the residual's variation can also accomodate a linear trend as is 
seen in the corresponding synthetic observations. We can only observe that the approximate 
magnitude of the residuals at early times are roughly the same for both real observations 
relative to the DE414 ephemeris and the synthetic observations with respect to their ephemeris. 
Thus, we cannot draw any conclusions about the existence or nonexistence of the Pioneer effect 
from our current knowledge of the orbit of Pluto; its observation arc is too short for such a 
determination. 

A third point with respect to Fig. \ref{fig14} is that post-discovery residuals are similar for both 
the synthetic observations and those of DE414 and very close for post-1960 observations. Overall, the 
only place where the two sets of normal points diverge is in the prediscovery era. Thus, neither set 
seems to indicate the existence or nonexistence of an additional gravitational anomaly. Both sets of 
residuals possess patterned irregularities, but it is impossible to say that one is superior to the 
other.

\begin{figure}
\plotone{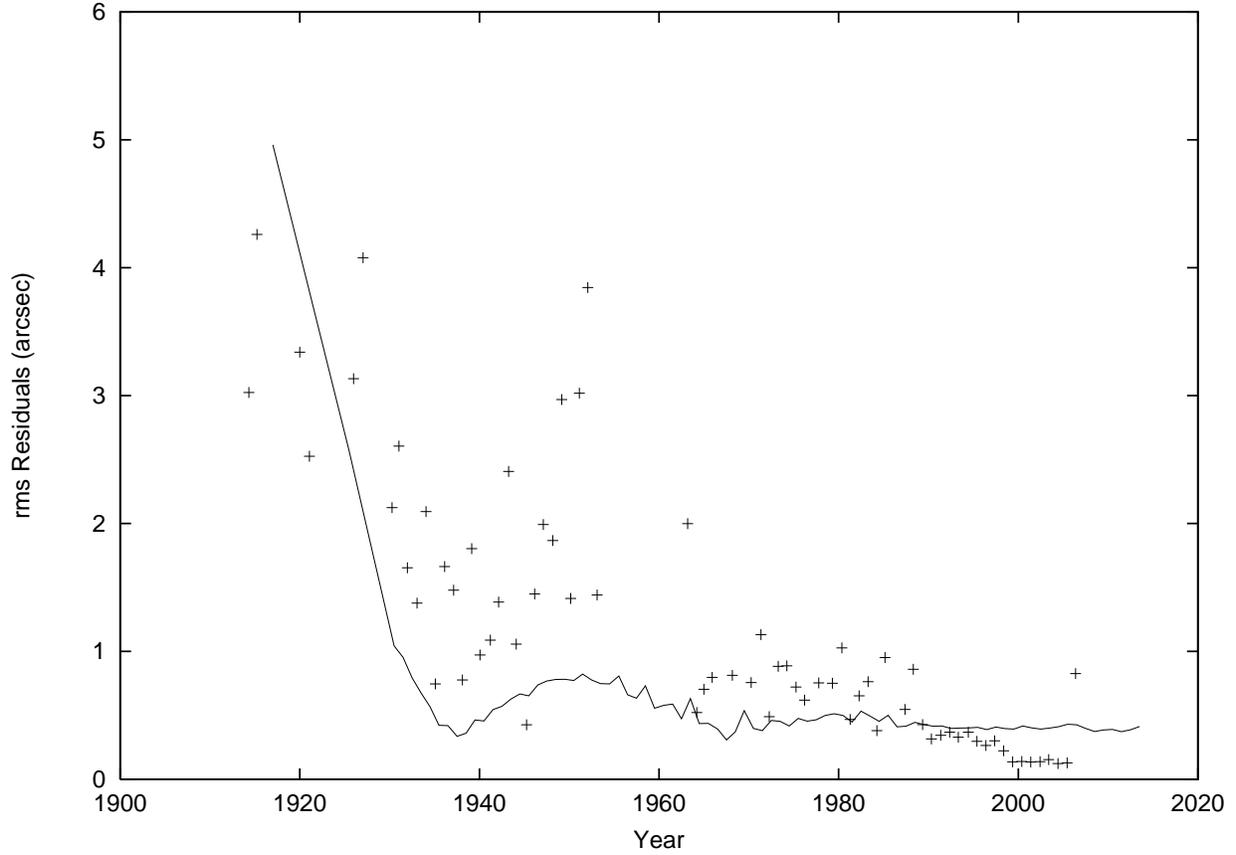}
\caption{\label{fig15}Total rms residual by epoch for the DE414 case and the synthetic observation 
case. (The total rms is the squre root of the sum of the squared residual in right ascension 
and declination, with an appropriate adjustment for the cosine of the declination.) The points 
show DE414's rms residual accumulated by opposition. The lines show the rms residual of the synthetic 
observations relative to their ephemeris accumulated by decade for prediscovery observations and by 
year thereafter.}
\end{figure}

As a final comment on the comparison of the two sets of residuals, Fig. \ref{fig15} shows the total 
rms residual by epoch for the two cases. This quantity is the square root of the sum of the squares 
of the residuals in right ascension and declination with an appropriate adjustment for the cosine of 
the declination. As above, DE414's residuals are accumulated by opposition, while the residuals of 
the synthetic observations relative to their ephemeris are accumulated by decade for prediscovery 
observations and by year thereafter. The quality of the residuals in the two cases is quite high 
over most of the observation arc. Even in the prediscovery era, the trends of the residuals are 
similar. As before, there is a linear trend in the synthetic residuals, but the entire prediscovery 
regime only includes two points. The trend of those two points is not dissimilar to that of the DE414 
residuals. Even if the prediscovery observations possessed a systematic error of as large as two 
seconds of arc, the trend in rms residual is not substantially different from those of the synthetic 
observations. 

From the discovery of Pluto to about 1960, the DE414 total rms residuals are greater than those found 
for the synthetic observations. This difference is expected due to the conservative assumptions of 
astrometric accuracy that were made in generating the synthetic observations. However, the trend of 
the residuals in both cases is similar, indicating a comparable fit to the observations. From about 
1960 onwards, both sets of residuals fit quite well. 

Given the likely, but uncharacterized, errors in pre-discovery observations, we are led 
to the observation that real observations are explained about as well by DE414 as the 
synthetic observations are explained by their ephemeris. 

The context of this comment must be kept in mind. The DE414 residuals relative to its ephemeris 
represent a fit to the actual observation archive by a normal gravity model without a Pioneer 
anomaly. The synthetic observations are created with a Pioneer anomaly present and thus their 
underlying motion is not in keeping with normal gravity. When the synthetic observations are fit 
to a gravity model without a Pioneer anomaly included in the dynamics, the fit is almost 
indistinguishable from that of the DE414 ephemeris. Especially given the uncharacterized errors 
apparently present in the prediscovery observation archive, we cannot distinguish between the two 
gravity models on the basis of the current observational record. However, we note that this is at 
least as much due to the shortness of the observation arc for Pluto as due to any difference in 
gravity models. 

What adds special interest to the Pluto case, however, is that due to its relatively large 
eccentricity Pluto shows potentially detectable differences between the two gravity models if 
the entire one hundred year observation arc were uniformly reliable. Thus, we are led again to 
the desirability of continuing to gather astrometric data on Pluto and expect that an astrometric 
difference in position might soon be able to falsify the existence of the Pioneer anomaly. 
Nevertheless, for the length of the currently available observation arc and because of the 
uneven quality of the observational archive, we cannot currently distinguish between the two 
gravity models on the basis of orbital position.

\subsection{Errors in orbital elements derived from observations}\label{sect3-B}

There are six orbital elements corresponding to the six degrees of freedom of the classical 
dynamical problem in three dimensions. Given that elements are determined from observations 
by a model fitting process, we must ask whether this large number of degrees of freedom is 
somehow redundant and whether a simpler model of the dynamics can be used to illuminate the 
problem.

This is the approach taken by \citet{2007PhRvD..76d2005T}, where synthetic observations are 
generated for a simplified dynamical model with four degrees of freedom. Conclusions are then 
drawn about the degree to which a Pioneer effect perturbation can be fit to this simplified 
model and thus whether the Pioneer effect exists.

The problem with this approach is that the elements that are ignored or suppressed in 
\citet{2007PhRvD..76d2005T} are precisely those with the greatest uncertainty resulting from 
the orbital fitting process. Further, as will be discussed below, care must be taken when 
using rms residuals as a primary measure of goodness of fit as far as model selection is 
concerned.

The underlying cause of the problems outlined above was discussed earlier. Briefly, this behavior 
is due to two inter-related factors: First, the orbital fitting problem is inherently nonlinear 
and is normally solved in the linear approximation; second, the sensitivity of the orbital 
solution to initial conditions is exacerbated, especially in the case of Pluto, by the problem 
of a short observation arc. These issues can sometimes be alleviated by a change in variables. 
Use of other than Keplerian orbital elements can result in an improved determination of orbital 
elements \citep{1961mcm..book.....B}. However, this type of variable change does nothing for the 
issues associated with a short observational arc. The uncertainties arising from these sources 
make the approach in \citet{2007PhRvD..76d2005T} problematic. In simplifying the problem by 
assuming away two or more degrees of freedom, the full parameter space, which has ample room to 
conceal the differences in predicted position, is not available for that purpose. This provides 
the false impression that perturbation effects are observable when, in fact, they are not.

One might think that the Pioneer acceleration, assumed to be radial, would only effect the 
motion of an orbiting body in its orbital plane. In a theoretical sense, this is true. The 
Pioneer effect exerts no torque on the orbiting body, and the plane of the orbit remains 
unchanged as time progresses. An analytical investigation of such an orbit can be considered 
as two dimensional and the number of degrees of freedom of the motion can be reduced accordingly. 
This is the essence of the approach of \citet{2007PhRvD..76d2005T}. However, in an observational 
context the motion is not restricted to a plane. There are at least two factors that force 
orbital motion to be considered to occur in three dimensional space with the associated degrees 
of freedom. The first is that unless the observer's location (e.g., Earth) remains in the plane 
of the orbit of the body, there will be parallax introduced which forces the motion into three 
dimensions. Similarly, observational errors will unavoidably make the object's position vary 
from its theoretical planar motion. Since both of these are a priori unknown and are only 
determined in the context of a least squares orbital solution for the observed object, the 
object's motion must necessarily be considered in its complete dimensionality. Of course, 
this requires all six degrees of freedom of the classical dynamical problem. There can be 
large errors associated with orbital elements and the extra degrees of freedom can provide 
ample room to ``hide'' orbital variations. This will be discussed in detail below. 

As shown in any numerical analysis text (for example, \citet[chapter 2]{heath1997}), an upper 
limit on the relative error of a fitted model parameter is related to the relative error of the 
independent variables by a condition number. The condition number can be estimated by the square 
root of the ratio of the largest to the smallest eigenvalues of the associated normal system of 
equations \citep[chapter 4]{heath1997}\footnote{\label{fn1}The differential correction problem 
can be stated as a linear matrix equation $Ax=b$, where $A$ is an appropriate Jacobian matrix, $x$ 
is a vector of elements, and $b$ is a vector of observations. The normal equations are formulated 
by multiplying this equation by the transpose of $A$ (denoted by $A^{T}$), giving $A^{T}Ax=A^{T}b$. 
This system of equations is solved formally by inverting the $A^{T}A$ product and multiplying again, 
giving $x=(A^{T}A)^{-1}A^{T}b$. In general, the $A$ matrix is not square. It has a column for each 
orbital element (e.g., six columns) and a row for each measured sky coordinate (e.g., twice the 
number of observations). However, the $A^{T}A$ matrix is square and possesses a corresponding set of 
eigenvalues. The ratio of the largest to the smallest eigenvalues of the $A^{T}A$ matrix provides 
the condition number of the normal equations. However, the condition number of the normal equations 
is the square of the condition number of the $A$ matrix. Thus, the original equation $Ax=b$ has a 
condition number equal to the square root of the condition number of the normal equations.}. 

It might be thought that the ill-posed nature of the orbital fitting problem could be obviated through 
a coordinate transformation. Indeed, there are alternative orbital elements that alleviate some 
difficulties with orbits of too small eccentricity or inclination, for example, equinoctial elements 
\citep{1961mcm..book.....B}. Alternatively, new variables that are linear combinations of the elements 
considered here can be used to alter the mix of errors among them. Thus, rather than specifying the 
time of perihelion passage, we can specify some instant of time and then specify either the mean 
anomaly, the mean longitude (the sum of the longitude of the ascending node, the argument of 
perihelion, and the mean anomaly), or the true longitude (the sum of the longitude of the ascending 
node, the argument of perihelion, and the true anomaly) at that moment. However, it can be shown that 
the condition number of a problem like this is invariant under linear coordinate transformations 
\citep[chapter 4]{heath1997}\footnote{\label{fn2}Matrices $A$ and $B$ are said to be similar when 
$B=C^{-1}AC$, where $C$ is nonsingular. Such transformations arise from changes of variables. Now, 
suppose we have an eigenvalue problem, $By={\lambda}y$ and let us perform this similarity 
transformation. Then $By={\lambda}y$ implies $C^{-1}ACy={\lambda}y$, which gives $A(Cy)={\lambda}(Cy)$ 
which can be stated as $Ax={\lambda}x$ so that $A$ and $B$ have the same eigenvalues and the 
eigenvectors of $A$ and $B$ are related by $x=Cy$. Thus, similarity transformations preserve 
eigenvalues but not eigenvectors. Since the condition number of a problem is the ratio of the largest 
to smallest eigenvalues, we see that condition number is preserved under linear coordinate 
transformations.} Although nonlinear coordinate transformations might alleviate this problem, there 
are none known to the authors that are used in practical astrometric work using OrbFit or other 
standard software packages. 

\begin{figure}
\plotone{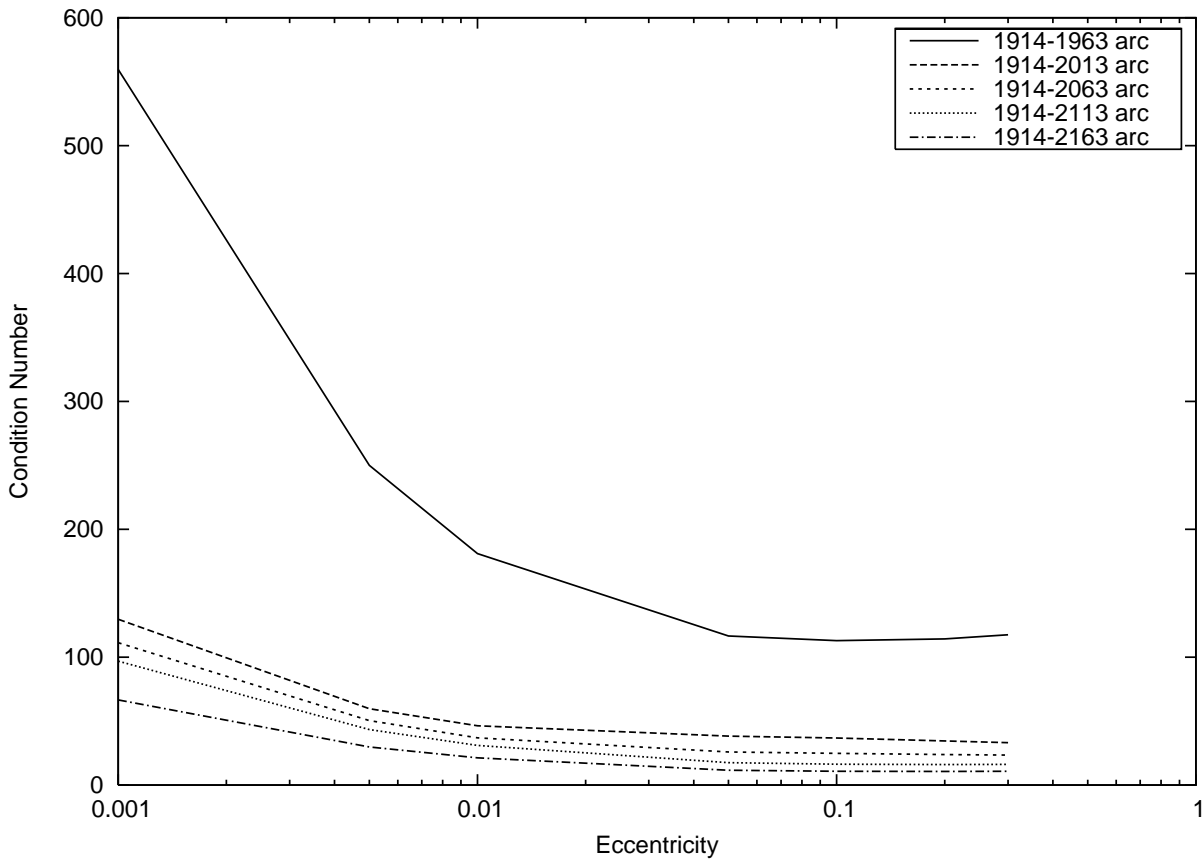}
\caption{\label{fig7}The condition number of the orbital fitting problem as a function of eccentricity 
for different observation arc lengths. For short arcs approximating one fifth of an orbital revolution, 
the condition number remains fairly large even for relatively large eccentricities. Low eccentricity 
objects, independent of arc length, always possess a relatively large condition number. These values 
of condition number are not large in the normal context of numerical analysis; however; the 
observationally driven orbital solution problem is much less precise than nominal machine precision. 
The ill-conditioning of the underlying mathematical problem coupled with the unavoidable observation 
errors can lead to large errors in the orbital elements.}
\end{figure}

Fig. \ref{fig7} shows the condition number of our orbital fitting problem as a function of orbital 
eccentricity and observation arc length. For low eccentricities, the condition number can become 
quite large. Even for eccentricities similar to that of Pluto, the condition number for a short arc 
problem implies large errors in orbital elements. 

Ordinarily, the condition number is important in the context of numerical roundoff with finite 
precision arithmetic in a computer. This is not the situation here; however, our precision constraint 
is the accuracy of the position observations that drive the orbital fitting process.

Thus an angular precision of 0.3 seconds of arc in right ascension and declination represents a 
relative error on the order of $10^{-4}$ percent or less. With a condition number of 100, this 
translates into an error of about 30 seconds of arc in orbital parameters like mean anomaly and 
argument of pericenter. 

\begin{figure*}
\plottwo{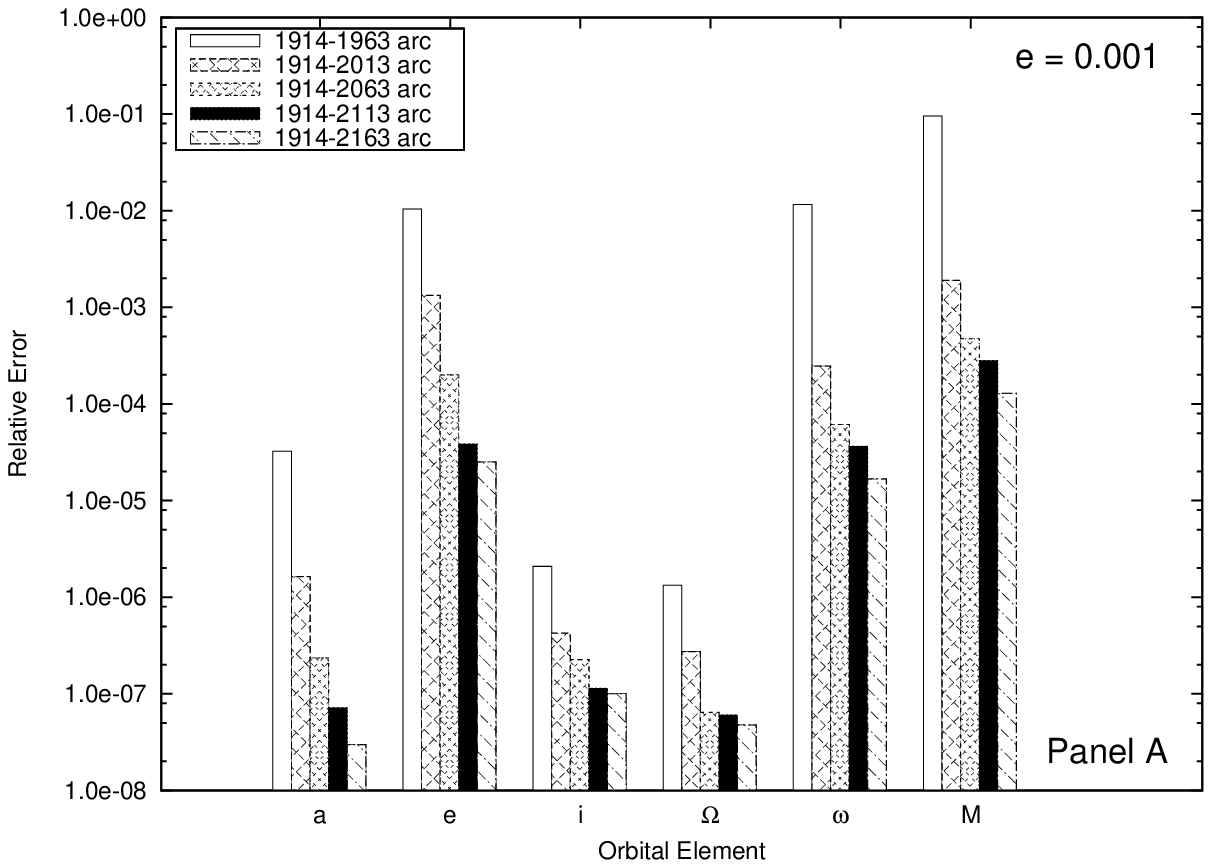}{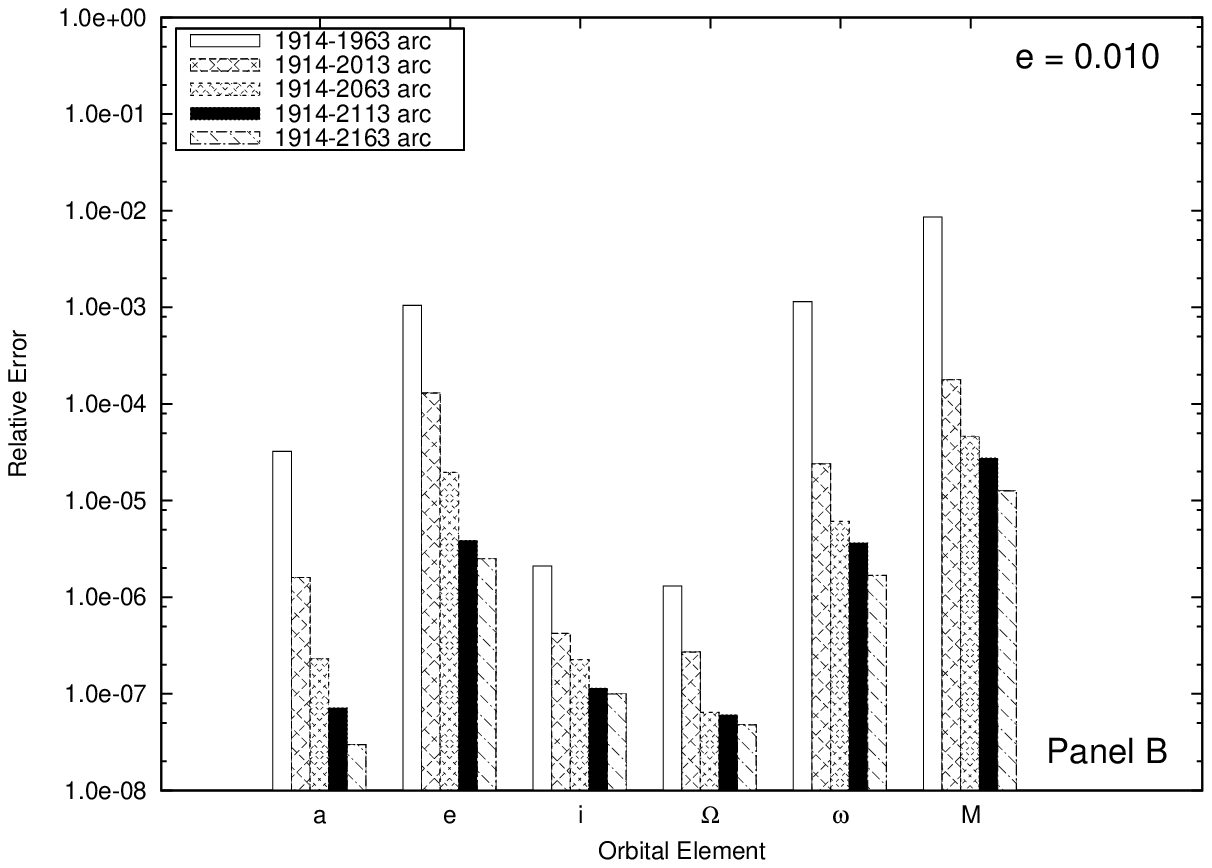}\\%
\plottwo{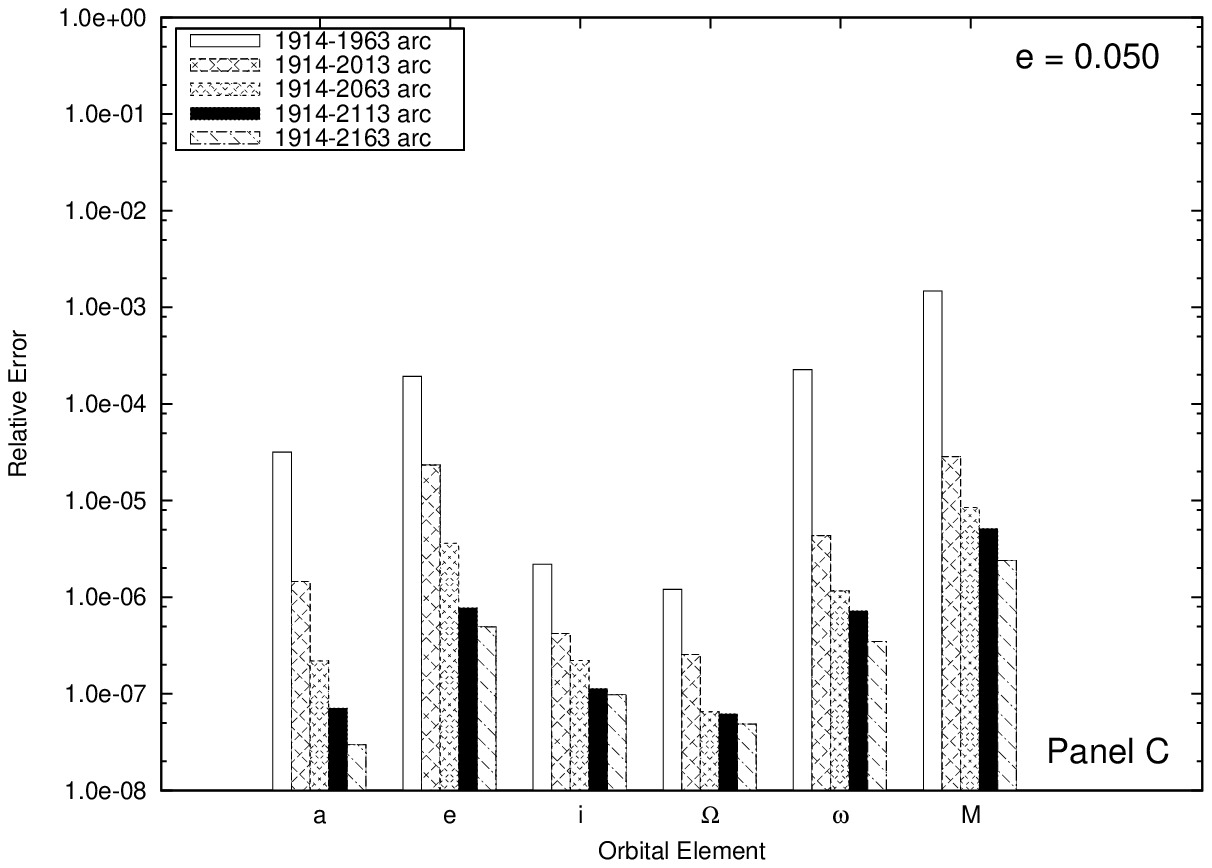}{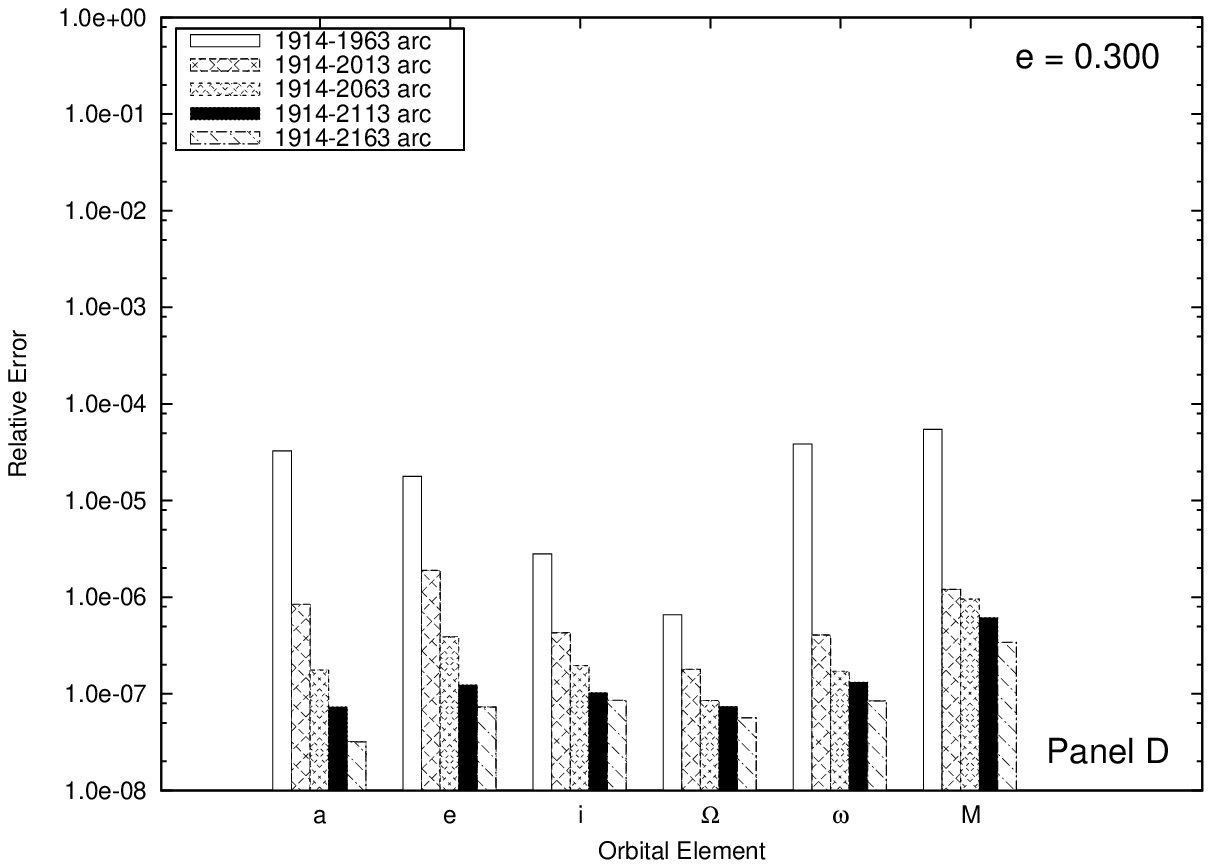}%
\caption{\label{fig8}Relative error in orbital elements in the linear approximation as observation arc varies. The cases 
shown are generated with a Pioneer perturbation present; these elements are all ``matched'' cases. Panels A through D 
show cases with progressively greater eccentricities. It is worth noting that a low-eccentricity object (Panel A) retains 
relative errors in argument of pericenter and mean anomaly of about one part in $10^{4}$ and $10^{5}$, respectively, even 
for a complete orbit. These errors amount to about four to five seconds of arc, respectively, over an order of magnitude 
greater than the nominal astrometric error. Thus, for a nearly circular orbit it is very difficult to differentiate 
between a Pioneer-perturbed and an unperturbed case. These errors drop progressively as the eccentricity increases. 
Panel D shows the corresponding case for a ``realistic'' Pluto, where the relative errors in these two elements are 
less than $10^{-6}$ for a complete orbit, amounting to about 0.05 arc second. Would that we had a complete orbit's 
observations for Pluto. Legend: $a$=semimajor axis; $e$=eccentricity; $i$=inclination; $\Omega$=longitude of ascending 
node; ${\omega}$=argument of pericenter; $M$=mean anomaly}
\end{figure*}

The corresponding results when we fit our synthetic observations bear out these theoretical 
considerations. Fig. \ref{fig8} shows the relative error (one sigma estimated error in the linear 
approximation) in orbital elements obtained with different observation arc lengths. The four panels 
show a representative set of orbital eccentricities. What is generally true for these figures is that 
for short arcs, the relative error in the elements is much greater than it is for the longer arcs. 
Similarly, the uncertainty in the location of the perihelion (as measured by the argument of the 
pericenter and the mean anomaly) is quite high and remains so for relatively long observation arcs. 
Additionally, since these errors are derived in the linear approximation, the real errors are likely 
to be greater, perhaps much greater. 

This implies a large error in mean anomaly or argument of pericenter, both associated with the spatial 
and temporal location of perihelion. This, coupled with the observational nature of the astrometric 
problem necessitating a full three dimensional treatment, forces us to conclude that a substantial 
amount of error can be absorbed into a multidimensional parameter space and, since it is unclear how 
the error can be allocated across the parameter space dimensions, the full parameter space must be 
included to properly reflect the uncertainties of the motion. Thus in order to drive the error in 
orbital elements down to levels allowing us to determine whether the Pioneer effect exists, we need 
a full dynamical model in three-dimensional space and either a relatively high eccentricity object 
or a long observation arc or both.

As observed before, our approach is itself an approximation. A ``full dynamical model'' should 
include not only adjustment of the orbital elements of Pluto, but also simultaneous adjustment 
to those of Uranus and Neptune as well. It is only in this way that all the second order 
perturbations to the system can be taken into account. However, as observed above, in order to 
illustrate the ideas concerned and the weaknesses of the approaches outlined above, we restrict 
our attention here to manipulating the orbit of Pluto. This approach will be justified further 
in Section \ref{disc}.

\subsection{\label{quality}How can we assess the quality of an orbital fit?} 

The problem of assessing model validity and comparing alternative models is huge and any kind of 
complete treatment is well beyond the scope of this paper. However, given the comments above, it 
is appropriate to offer some observations on this large and interesting topic.

The issue of goodness-of-fit for models is not a simple one. Often, a $\chi^2$ is calculated and 
compared to determine which of competing models is the preferred. However, there are complexities 
involved. For example, \citet{2003sppp.conf...70N} discusses goodness-of-fit and points out that 
while use of the $\chi^{2}$ statistic as a goodness-of-fit measure for binned data is justifiable 
and is often done, it has flaws. He further observes that for unbinned data (such as we have here), 
there is no equivalent popular method for measuring goodness-of-fit. Indeed, \citet{2003sppp.conf...52H} 
gives several examples of problematic goodness-of-fit cases. Similarly, \citet{NIST_EngrStatHB} 
advocates investigating the structure of residuals to find patterns, biases, and sytematic 
differences between the model and the data. 

\begin{figure*}
\plottwo{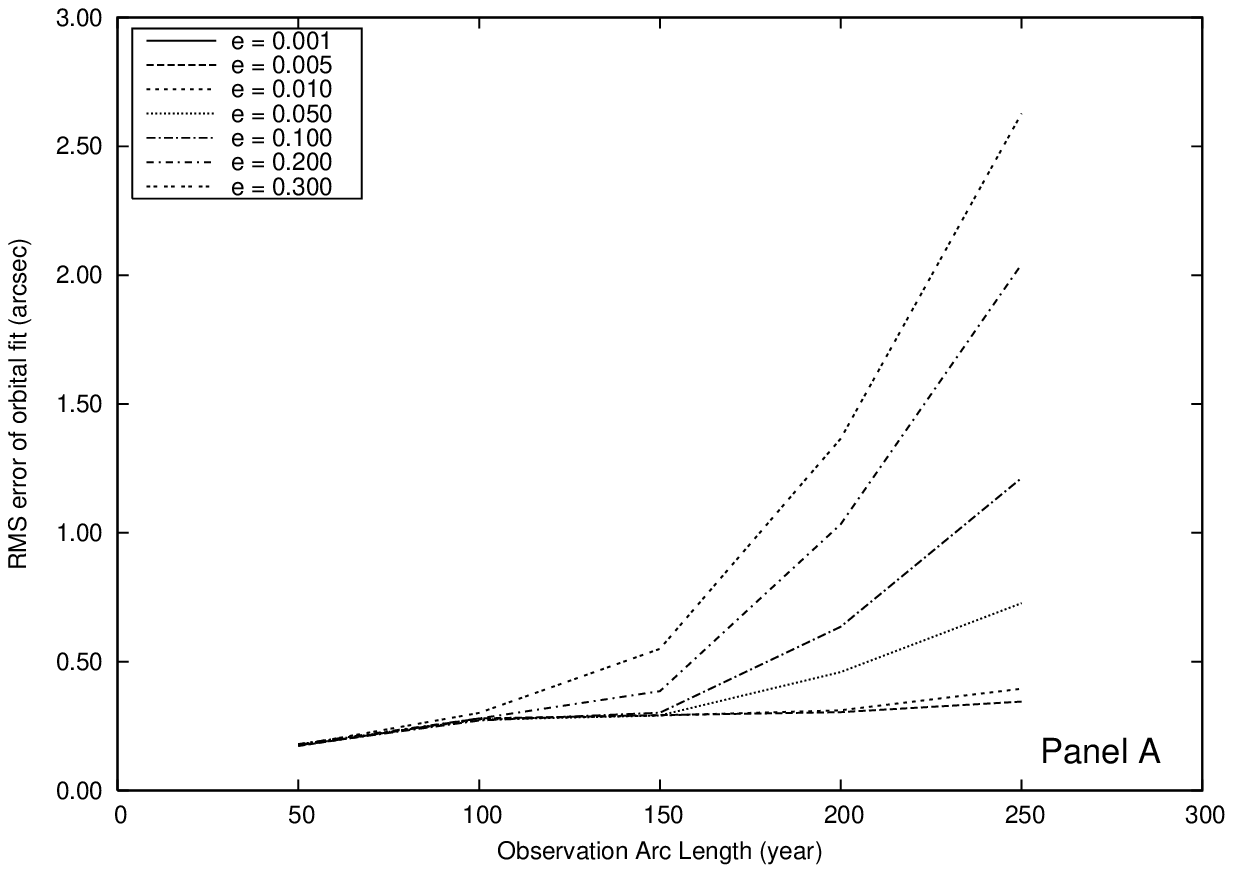}{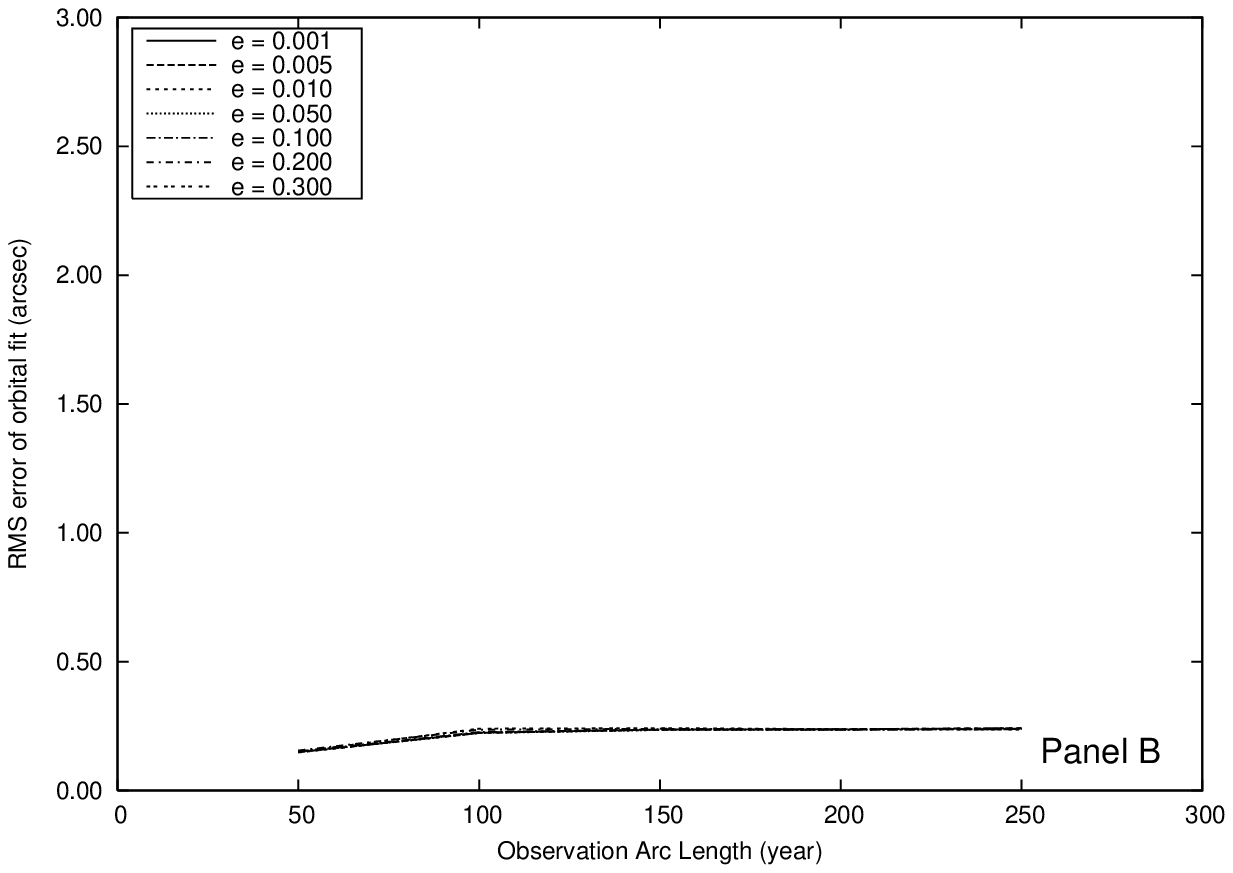}%
\caption{\label{fig9}Rms residual of orbital fit as observation arc length varies. Observations are generated 
with a Pioneer effect present. Panel A shows the ``mismatched'' case where the observations are fit to a gravity 
model not containing the Pioneer effect. Panel B shows the ``matched'' case where the same observations are fit 
to a gravity model that includes a Pioneer effect perturbation. Note that for observation arcs up to the present, 
all with about 100 year observation arcs, there are only slight differences between the two gravity models in 
terms of their quality of fit as measured by the rms residual (see Fig. \ref{fig3}).}
\end{figure*}

We can divide metrics for model quality into two broad categories that, although useful, are certainly 
not mutually exclusive. First, we can consider what might be termed point estimates of fit quality. 
Often, a measure of merit is used that is related to the rms residuals between the model and the data. 
Fig. \ref{fig9} compares our rms residuals for various eccentricities. One panel of this figure shows 
the results of fitting the ``mismatched'' case of the Pioneer-perturbed observations being fit to a 
strictly Newtonian gravity model; the other panel shows the corresponding ``matched'' case where the 
gravity model includes a Pioneer perturbation. In both cases the underlying synthetic observations 
contain a Pioneer perturbation. The most striking feature of this figure is how slowly the residual 
degrades as the arc length increases, especially for low eccentricity objects. Indeed, until the 
observation arc is over 100 years long, there are only small differences in the quality of the fit 
as measured by rms residuals for any of the eccentricities evaluated. Even in the ``mismatched'' 
case, the rms residual is still less than half an arc second when the arc is 150 years long. In 
keeping with the discussion above about the ill-conditioned nature of the orbital fitting problem, 
we see that strictly relying on rms residuals as a measure of merit for model selection can be 
problematic.

Another measure of merit that is sometimes invoked is the uncertainties in the orbital elements 
themselves. The orbital fitting process automatically provides mathematically well-justified 
estimates of the elements, although these are in a linear approximation that has its own difficulties 
since the orbital fitting problem is intrinsically nonlinear. This approximation provides error 
estimates similar to those shown in Fig. \ref{fig8}. Across the eccentricities shown, there is 
substantial error spread throughout the parameter space. In particular, it might be instructive 
to compare Panels A and D in Fig. \ref{fig8}. In Panel A, a very low eccentricity case, the errors 
in semimajor axis, inclination, and longitude of the ascending node drop to a level allowing seven 
or eight significant digits to be present in the corresponding element values. At the same time, 
however, the uncertainties in eccentricity, argument of the pericenter, and mean anomaly remain 
relatively large. In the case with the largest eccentricity evaluated, Panel D shows the uncertainty 
to be more uniformly spread across elements. Trading off the uncertainties of one element against 
those of another to compare two models is therefore at best arbitrary and could be misleading.

Position error in the sky plane is yet another point metric for fit quality. In this case, 
uncertainties in the orbital elements are mapped directly onto the sky. Here, all the comments 
above about the uncertainty of the elements are valid, plus the observation that the mapping 
from the six-dimensional space of orbital elements to the two-dimensional sky plane is highly 
degenerate; many sets of elements can map to the same region of sky. Thus, the sky plane position, 
by itself, can have problems as a point measure of model fit quality. In particular, comparing 
the nominal solution for two models can result in a large error volume about the calculated 
positions of objects.

All three approaches to providing point estimates of model fit quality have one common weakness. 
They are all based on a linearized form of the orbital fitting problem and make the assumptions 
associated with least squares fitting like normally distributed errors, independence, and no bias 
or systematic errors. Possibly the biggest manifestation of this is that we should not expect to 
be able to accurately extrapolate very far beyond the available observation arc unless there is 
at least a full orbital revolution's volume of data and even then, extrapolation is dangerous. 

A possibly better approach to comparatively assessing model alternatives is to use some type of 
global comparison. Of broadest applicability in this context is the idea of testing for the 
normality of the residuals between the model and the data. The overall basis for these tests is 
to look at the structure of the residuals and to perform statistical tests for lack of fit (see, 
for example, \citet[Section 4.4.4.6]{NIST_EngrStatHB}). The basis for these tests is to search 
for patterns, biases, and systematic differences between the model and the data. 

Another approach that could be used to address the global quality of fit is to use bootstrap 
techniques. This approach could involve, as done here, synthesizing observations whose error 
characteristics reflect those found actual observations and then using the synthetic observations 
to assess actual variations in the fit quantities. Another approach of this sort would be to 
remove some of the data from the fit and see how the fitted solution extrapolates to the times 
that were removed. A similar approach would be simply to remove data randomly from the observation 
set and analyze the resulting variations in fitting parameters. 

One form that this approach could take would be to conduct a full Bayesian analysis of competing 
models. The distributions of orbital elements resulting from synthetic observations that arise 
from a monte carlo process could be used to compare the probability of obtaining the observations 
in light of the competing models (see, for example, \citet{Jaynes2003, Gregory2005, Sivia2006}). 
Although beyond the scope of this paper, the findings discussed here show that there are not 
currently enough data to warrant this type of analysis; however, the advent of Pan-STARRS and 
LSST will change that situation in the near future.

\section{\label{disc}DISCUSSION}

In the analysis described above we first showed that one cannot simply take orbital elements 
resulting from the fit of observations to a particular force model and use them to predict 
positions resulting from motion under the influence of another force. Rather, we must refit 
the orbits to the observations under the new force model. In the case of Pluto, this produces 
a well-fitting orbit that is indistinguishable in a practical sense, at least as long as the 
observation arc is short enough, from the unperturbed motion. Thus, without redetermining the 
elements we cannot make sweeping generalizations about whether or not the outer planets' orbits 
show that the Pioneer effect does or does not exist.

Similarly, although making simplifying assumptions about a physical situation in order to draw 
conclusions is a time-honored theoretical mode of attack, if the physical model is oversimplified 
we can be misled into erroneous conclusions. As seen above, in an observational context, both 
observer position and observational errors lead to the necessity of introducing the third 
spatial dimension with its associated degrees of freedom. If we do not keep an appropriate number 
of degrees of freedom, the problem can be oversimplified too much and mislead us into unwarranted 
conclusions. In particular, we must take care in using such a simplified model to conclude that 
the Pioneer effect does not exist.

\begin{figure}
\plotone{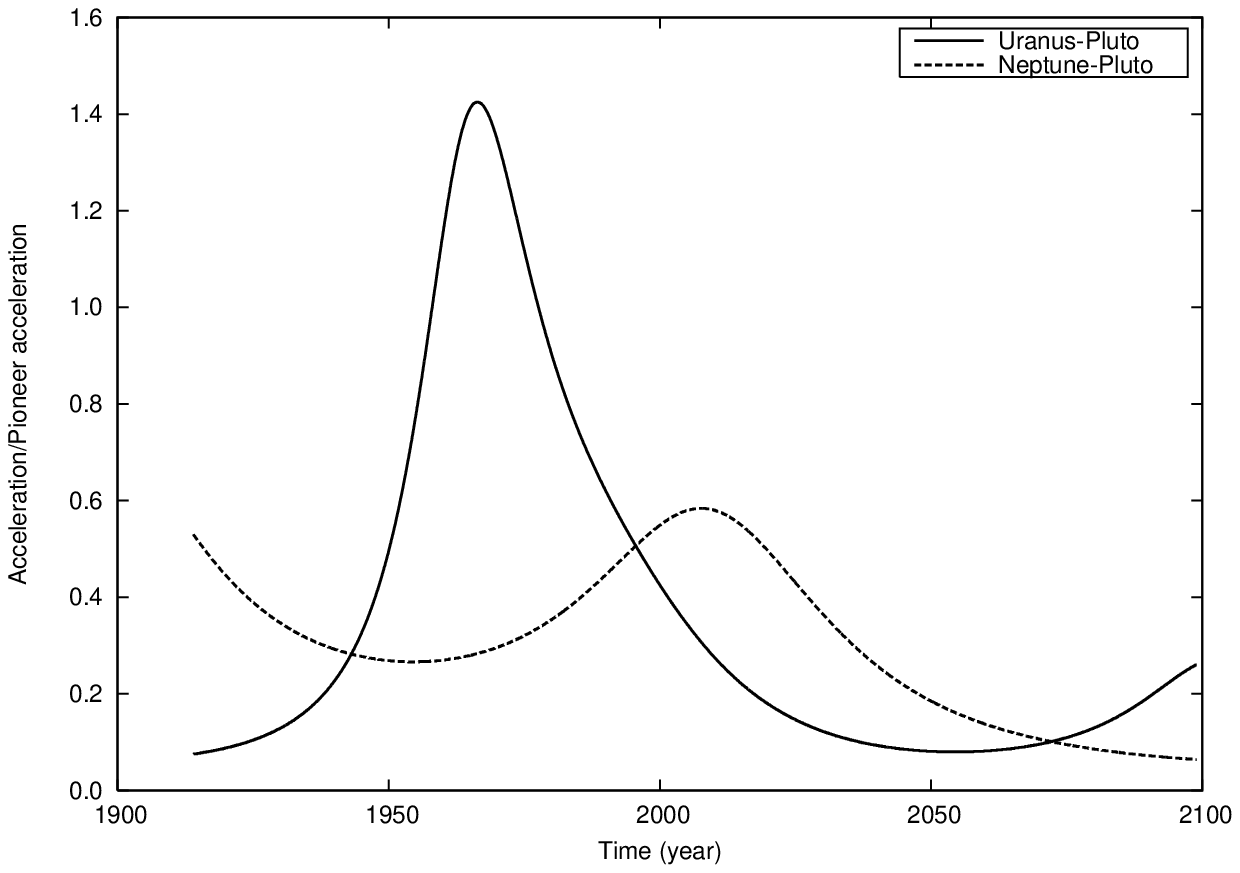}
\caption{\label{fig16}Gravitational acceleration exerted by Uranus and Neptune on Pluto as a 
function of time. The acceleration is shown in units of the Pioneer acceleration. Both are of 
the order of the Pioneer acceleration. While one might think that Neptune's influence on Pluto 
would be greater than that of Uranus, the reverse is true due to the relative positions of the 
three planets in the solar system.}
\end{figure}

Once again, it should be emphasized that our approach is itself an approximation. To properly 
conduct an analysis of the sort outlined here, the orbital parameters of the entire system 
of outer planets should be included. This would bring into the calculation all second order 
perturbations. However, we have taken a simplified approach and believe it accurately and 
fairly addresses the uncertainties in the orbit of Pluto. One reason we believe our simplified 
approach is valid is illustrated in Fig. \ref{fig16}.

Fig. \ref{fig16} shows the force per unit mass (in units of the Pioneer acceleration) exerted 
by Uranus and Neptune at Pluto's position over the period of time we have considered. Perhaps 
surprisingly, due to the relative positions of the planets in their orbits, Neptune exerts 
less force on Pluto than Uranus over most of this period. The magnitude of the force is of 
the order of the Pioneer acceleration for both planets. If the Pioneer effect exists, we would 
expect the orbital elements of Uranus and Neptune to change, but their positions would change by 
very little. The magnitude of the forces they exert on Pluto would therefore change by an amount 
much less than the current magnitude of those forces. Thus, we argue that the approach used in this 
paper, while not as accurate as a full second order calculation involving all the outer planets, is 
accurate enough for our purposes. 

Given the comments above, how then are we to compare alternative models? There are a number of 
methods that we can use to compare and assess models and their results. In the orbit fitting 
context, these have been discussed above in terms of point estimates and global estimates of 
goodness of fit. The simplest point estimate is to simply calculate the rms residual of the 
fit compared with the input observations. However, as we've seen, the ill-conditioned nature 
of the orbital fitting problem can sometimes make the residual a poor candidate for this role. 
Errors, in the linear approximation, to the orbital elements can be assessed to determine the 
quality of the fit. Further physical insight can be gathered from inspecting the sky position 
errors that result from element errors. Global assessments of model fit primarily revolve around 
the normality of residuals. There are lack of fit tests that can be used to test the residuals 
and determine if there are any indications of a deficient model. Generally, a lack of fit is 
manifested as patterns, biases, and systematic variations in the residuals, which would indicate 
a poorly fitting model. 

Which of these metrics is best? The reality is that determining the quality of a model and comparing 
model effectiveness is a complex problem. We need all these measures of model fit. Statistical tests 
should be performed comparing predicted positions, taking proper account of the associated errors, 
to test the hypothesis that the positions predicted under the two dynamical models are different. 
Only in this way can the existence of a perturbation like the Pioneer effect be falsified through 
astrometric methods. 

As discussed earlier, Pluto is very interesting in that orbital fits to the two gravity models 
show deviations in predicted position that are probably detectable if the entire observation arc 
consisted of reliable measurements. However, prior to about 1960, there is a significantly greater 
dispersion in residuals that at later times. Only about half of Pluto's motion since its discovery 
has been the result of systematic and organized observing campaigns \citep{1994Icar..108..174G}. 
What adds special interest, however, is that because of its relatively large eccentricity Pluto 
is likely to show differences in position predictions for the two gravity models in the relatively 
near future as more observations are accumulated. 

\section{\label{concl}CONCLUSIONS}

The analysis described above shows two major things. First, we must fit observations to a 
particular dynamical model and adjust orbital elements before predicted positions on the sky 
can be compared. Orbital parameters are derived from observations which have associated an 
unavoidable error. The determination of orbits is a model fitting process which has its own 
associated error sources. Extrapolating sky positions very far past the end of an observation 
arc can result in predicted observations becoming inaccurate so rapidly as to be worthless. 
The implication of these findings is that ``matched'' and ``mismatched'' gravity models 
cannot be distinguished on the basis of observable sky positions for observation arc lengths 
similar to those currently obtaining for Pluto. 

Similarly, in order to draw conclusions about differences in position in the sky, we must be 
careful not to oversimplify the dynamical model used to draw the conclusions. Suppressing 
degrees of freedom in the dynamics simplifies the orbital determination problem to just such 
a degree. The orbital determination problem is nonlinear and the customary solution methods 
are approximations. Thus, any missing or ignored degrees of freedom can, if present, conceal 
dynamical effects associated with differing gravity models; we are forced to make use of the 
full dimensionality of the dynamical problem. In particular, a substantial amount of variation 
can be absorbed into a multidimensional parameter space and the full parameter space must be 
considered to properly reflect differences in motion of the outer planets due to the Pioneer 
effect.

The problem with the simplified approach is a two-headed one. First, the orbital fitting problem 
is inherently nonlinear and is normally solved in the linear approximation. Even if not 
mathematically chaotic, the system of equations is sensitively dependent upon initial conditions. 
Thus, small changes in elements can result in large changes in predicted position outside the 
range of observations. This sensitivity is exacerbated by the problem of a short observation arc. 
The length of the entire observational archive for Pluto is less than about one-third of a 
complete revolution. Together, these factors conspire to potentially generate large errors 
outside the observation arc, while increasing the length of the observation arc can markedly 
reduce error over the whole of the arc and even beyond it. 

We must conclude that we do not know the orbit of Pluto as well as we might have thought. We must 
continue to perform astrometry on it in order to be able to comment on the accuracy with which we 
know its orbit. Using current data, we cannot assert that the motion of Pluto demonstrates that 
the Pioneer effect does not exist. That jury is still out. Of course, this does not mean that the 
Pioneer effect exists. It does mean that we cannot deny the existence of the Pioneer effect on the 
basis of motions of the Pluto as currently known. Further observations are required before such an 
assertion can be made with confidence.

It should again be emphasized that our approach is itself an approximation. The dynamical 
system that should be analyzed to provide a comprehensive answer to the question of the 
detectabilty of the Pioneer effect should include not only adjustment of the orbital elements 
of Pluto, but also simultaneous adjustment to those of Uranus and Neptune as well. It is 
only in this way that all the second order perturbations to the system can be taken into 
account. Our approach here, however serves to illustrate the ideas concerned and the 
weaknesses of the approaches outlined above. 

Finally, it should be pointed out that, in addition to the observations of individual 
objects discussed in this paper, there are other related approaches to assessing gravity in 
the outer solar system. For example, recently \citet{2007ApJ...666.1296W} have investigated 
the use of ensembles of Trans-Neptunian Objects (TNOs) to ascertain whether their motion 
reflects unknown additional perturbations and showed that the Pioneer effect was not consistent 
with the motion of TNOs. On the other hand, in the area of the observation of individual 
objects as discussed in this paper, the advent of Pan-STARRS and LSST in the next several years 
should provide sufficient data to determine whether the motion of outer solar system bodies 
reflect the action of unknown forces. This determination should occur over time frames 
discussed in this paper. However, a combination of the techniques of \citet{2007ApJ...666.1296W} 
and the considerations presented here should provide definitive answers more quickly.

\acknowledgements

The authors wish to thank Dr. Myles Standish for providing the residuals of JPL's DE414 
ephemeris. We also wish to thank the anonymous referee for a close reading of the manuscript 
and comments that resulted in significant improvements to the paper. 

%\bibliography{GLP_bibliography}   % added by GLP 10-23-08

\end{document}